\documentclass[fleqn,usenatbib,useAMS]{mnras}

\usepackage{graphicx}	
\usepackage{amsmath}	
\usepackage{amssymb}	
\usepackage{multicol}        
\usepackage{diagbox}
\usepackage{bm}		
\usepackage{pdflscape}	

\newcommand{\kmps}{km~s\ensuremath{^{-1}}\,}

\newcommand{\Msunyr}{M\ensuremath{_\odot}~yr\ensuremath{^{-1}}\,}

\def\kms{$\textrm{km~s$^{-1}$}$}

\DeclareRobustCommand{\ion}[2]{%
	\relax\ifmmode
	\ifx\testbx\f
	{\mathrm{#1\,\textsc{#2}}}\else
	{\mathrm{#1\,\mathsc{#2}}}\fi
	\else\textup{#1\,{\mdseries\textsc{#2}}}%
	\fi}
\newcommand{\MC}{\multicolumn}

\usepackage[T1]{fontenc}
\usepackage{ae,aecompl}
\usepackage{newtxtext,newtxmath}
\usepackage{epstopdf}

\title[Rings in NGC 1533 and NGC 1543]{Spectral study of starforming rings in S0 galaxies of Dorado group -- NGC 1533 and NGC 1543}
\author[I. Proshina et al.]{
Irina S. Proshina,$^{1}$\thanks{E-mail:ipro@sai.msu.ru}
Olga K. Sil'chenko,$^{1}$\thanks{E-mail:olga@sai.msu.su}
Alexei Yu. Kniazev$^{2,3,1}$ \\
$^1$ Sternberg Astronomical Institute, Moscow M.V. Lomonosov State University, Universitetskij pr., 13,  Moscow, 119234, Russia\\
$^2$ South African Astronomical Observatory, P.O. Box 9, Observatory, Cape Town, 7935, South Africa \\
$^3$ Southern African Large Telescope, P.O. Box 9, Observatory, Cape Town, 7935, South Africa  \\
}

\begin{document}
\label{firstpage}
\pagerange{\pageref{firstpage}--\pageref{lastpage}} \pubyear{2023}
\maketitle

\begin{abstract}
We have fulfilled a detailed long-slit spectroscopic analysis for two SB0 galaxies -- NGC~1533 and NGC~1543, -- 
belonging to the Dorado group. Our spectral data reveal asymmetric decoupled kinematics of the stars and ionised gas
in these barred lenticular galaxies that give evidences for external origin of the gas in the rings. We have
calculated the star formation rates in the rings by using the ultraviolet fluxes of the rings corrected for
the foreground and intrinsic absorption; and we have estimated parameters of the stellar populations in the
inner parts of the galaxies confirming that they are old -- except the nucleus of NGC~1543, which demonstrates
signatures of re-juvenation less than 5~Gyr ago.
\end{abstract}

\section{Introduction}\label{intro}

Lenticular galaxies were introduced by \citet{hubble} as disc galaxies without star formation locating
at the position intermediate between ellipticals and spirals at the famous Hubble's fork which described his morphological
classification scheme. Since global stellar structures of spirals and lenticulars were similar, consisting of
centrally-concentrated bulges and extended flat discs, the absence of star formation in lenticulars was initially
explained by their gas deficiency \citep[e.g.][]{biermann_tinsley}. However, later deep radio surveys have
revealed the presence of cold gas in the most lenticulars \citep{welch2010}. Half of gas-rich lenticulars
demonstrate also weak star formation \citep{pogge_eskridge}, and this star formation is usually organised into
ring structures \citep{thilker2007,salim12}. The neutral-hydrogen discs in massive S0s are very
extended \citep{Oosterloo2007}, and the star formation in S0s is often proceeding in the outer low
surface-brightness discs: among the lenticulars with outer current star formation studied by us recently in detail
we can refer to UGC~4599 \citep{u4599} or to NGC~934 \citep{n934}.
Under the popular now concept that the evolution of disc galaxies is governed by persistent cold-gas accretion, a suggestive hypothesis
can be formulated that the difference between lenticulars and spirals may be due to different regimes of gas accretion. 
In particular, the large extension of their gaseous discs may be an evidence in favour of higher angular momentum
of the accreted gas in S0s \citep{Peng_Renzini}. Or high frequency of decoupled gas-star rotations in S0s may be the evidence
in favour of inclined orientation of gas inflows resulting in gas heating and star formation suppression \citep{s0_fp}.

\begin{figure*}
\begin{tabular}{c c}
 \includegraphics[width=0.48\textwidth]{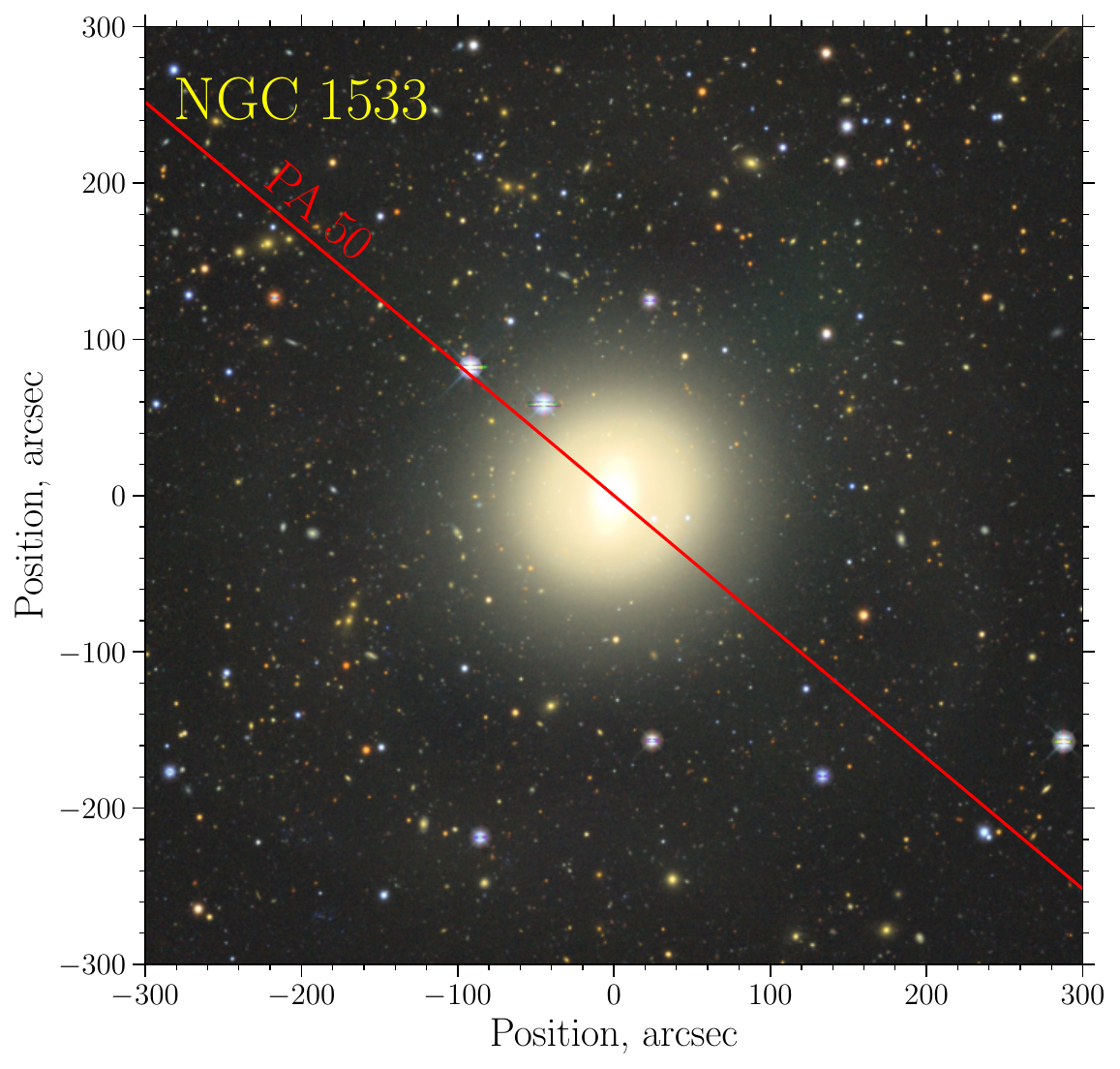} &
 \includegraphics[width=0.48\textwidth]{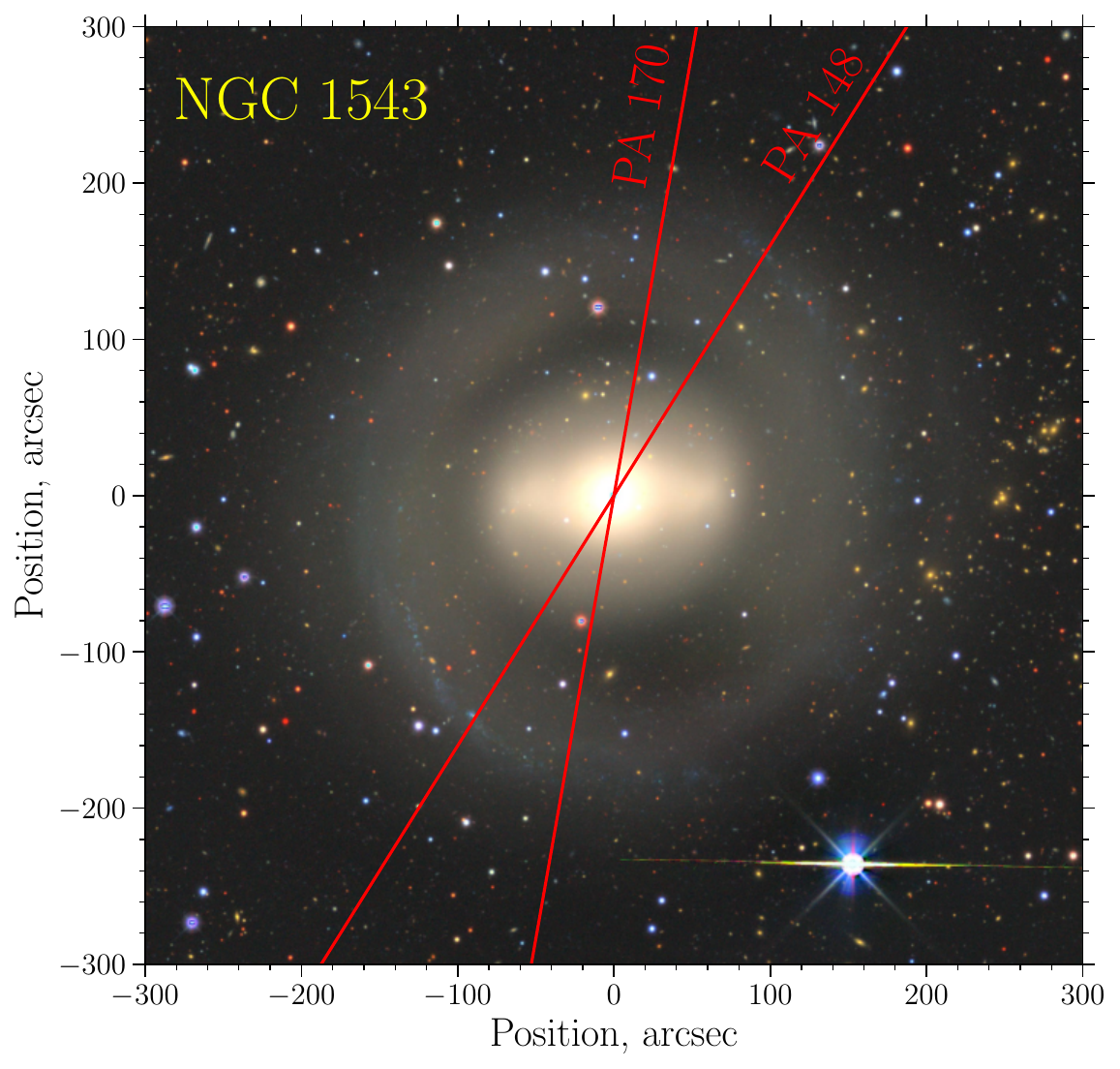} \\
\end{tabular}
\caption{The full-colour images of the galaxies studied -- NGC~1533 at the {\it left plot} and NGC~1543
at the {\it right plot}, with the spectrograph slit orientations overposed. The images are taken from the
Legacy Survey resource \citep{legacysurvey}.
}\label{images}
\end{figure*}

Since the evolution of lenticular galaxies may be defined by their nearest environments, it is also important to
characterize in detail the galaxy environments when trying to identify their evolutionary paths. In this work we
study two intermediate-mass lenticular galaxies, having their spatial orientation close to face-on, which have large ultraviolet 
rings thus experiencing star formation in their outermost regions. These galaxies are NGC~1533 and NGC~1543 belonging to the 
nearby rich galaxy group Dorado. The galaxy images are presented in Fig.~\ref{images}, and their main characteristics
are given in Table~\ref{tbl_lit}.

\begin{table}
\caption{Global parameters of the galaxies under consideration}
\label{tbl_lit}
\begin{flushleft}
\begin{tabular}{l|cc}
\hline\noalign{\smallskip}
Galaxy & NGC~1533 & NGC~1543  \\
\hline
Type (NED$^1$) & (L)SB(rs)$0^0$ & (R)SB(s)$0^0$ \\
$V_r $ (NED), $\mbox{km} \cdot \mbox{s}^{-1}$ &  790 & 1176 \\
$R_{25}^{\prime \prime}$(RC3$^2$) & 83 & 147 \\
$R_{25}$, kpc & 8.35 & 13.4  \\
$M_B$(LEDA$^3$)  & --19.72$^m$ & --20.14$^m$ \\
$M_H$(NED)  & --23.63$^m$ & --23.45$^m$ \\
Distance$^4$, Mpc  & 20.89 & 18.71 \\
Linear scale, kpc per $^{\prime \prime}$ & 0.101 & 0.091  \\
$PA(phot)^5$ & $120^{\circ}\pm 5^{\circ}$ & $9^{\circ}\pm 6^{\circ}$ \\
$i(phot)^5$ & $22^{\circ}$ & $18.6^{\circ}$ \\
$ \lg (M(\mbox{HI}) ^6, M_{\odot}) $ &  $9.80\pm 0.08$ & $8.92\pm 0.07$ \\
$ \lg (M_* ,  ^6, M_{\odot}) $ &  10.60 & 10.54 \\
$\Delta ^7$, kpc (NED) &  344  & 660 \\
\hline
\multicolumn{3}{l}{$^1$\rule{0pt}{11pt}\footnotesize
NASA/IPAC Extragalactic Database (http://ned.ipac.caltech.edu).}\\
\multicolumn{3}{l}{$^2$\rule{0pt}{11pt}\footnotesize
Third Reference Catalogue of Bright Galaxies \citep{rc3}.}\\
\multicolumn{3}{l}{$^3$\rule{0pt}{11pt}\footnotesize
Lyon-Meudon Extragalactic Database, \citet{hyperleda}. }\\
\multicolumn{3}{l}{$^4$\rule{0pt}{11pt}\footnotesize
Cosmicflows-3 \citep{Courtois2013,Cosmicflows3}} \\
\multicolumn{3}{l}{$^5$\rule{0pt}{11pt}\footnotesize
Carnegie-Irvine photometric survey, \citet{Carnegie-Irvine}.} \\
\multicolumn{3}{l}{$^6$\rule{0pt}{11pt}\footnotesize
\citet{Murugeshan2023}.}\\ 
\multicolumn{3}{l}{$^7$\rule{0pt}{11pt}\footnotesize
projected separation in kpc from the group center (NGC~1553).}\\
\end{tabular}
\end{flushleft}
\end{table}

The nearby galaxy group, Dorado group, or Dorado cloud, consists of a dozen galaxies, mostly early-type ones \citep{Tully_groups},
and in particular NGC~1533 and NGC~1543 belong to the red sequence \citep{Cattapan2019}. The
group is loose; as for its dynamical status, it is unvirialized \citep{Firth2006}. After mapping the
large-scale structure in the nearby Universe, \citet{Shaya2017} have identified a filament going
through the Dorado group, Milky Way, and Antlia cluster. Peculiar velocities along this filament
are rather large, so the distances determined through the Hubble law are unreliable in this case.
We use the distances to NGC~1533 and NGC~1543 derived in the Cosmicflows-3 project \citep{Courtois2013,Cosmicflows3} 
by taking into account peculiar velocities; just these distances are given in Table~\ref{tbl_lit}.
By fixing the spatial positions of NGC~1533 and NGC~1543 we find that they are at the very periphery
of the group: the projected separation between NGC~1543 and the center of the group (NGC~1553) is 660~kpc
and the line-of-sight separation is 1.2~Mpc while NGC~1533 is projected closer to the center, in
344~kpc, but the line-of-sight separation is 3~Mpc. We conclude that the galaxies are located in very
rarified field. NGC~1543 looks quite isolated, while NGC~1533 has two close dwarf satellites, IC~2038 and
IC~2039, in some 60~kpc from it.

Ultraviolet (UV) features of recent star formation were found in NGC~1533 and NGC~1543 long ago. Over this field great 
efforts were applied by Roberto Rampazzo with co-authors. Firstly, they have used UV data of the spacecraft
Swift \citep{Trinchieri_Swift,Rampazzo_Swift} to discover outer UV-bright rings in these SB0 galaxies --
with the radius of up to $R_{ring}\sim 50^{\prime \prime}$ in NGC~1533 and up to $R_{ring}\sim 3^{\prime}$
in NGC~1543. Secondly, narrow-band photometry was undertaken, with the filter including the emission-line
complex H$\alpha +$[NII]$\lambda$6583, and the ring structures of NGC~1533 and NGC~1543 were traced with
another indicator of recent star formation; the integrated star formation rates (SFR) were estimated as 
0.4\Msunyr per every galaxy, NGC~1533 and NGC~1543 \citep{Rampazzo2020}. The cold gas reservoirs feeding
this star formation were also found -- by \citet{RyanWeber2004} in NGC~1533 and by \citet{Murugeshan2019}
in NGC~1543, in the latter galaxy the HI ring being coincident with the UV-ring. Finally, NGC~1533 has been
deeply mapped with the Indian UV space telescope Astrosat/UVIT, and an outer low surface brightness UV ring
was found which is coincident with the HI distribution, rather peculiar around NGC~1533. The outer weak star
formation in NGC~1533 and NGC~1543, feeding by the extended gas reservoirs, is now firmly established.

The information which is still deficient for the starforming structures in the SB0 galaxies NGC~1533 and
NGC~1543 is spectral information. Meantime it is quite necessary to refine our knowledge about the exact
nature of gas excitation in the ring, about the gas rotation, and kinematics in general, and about the gas
chemistry. We have undertaken long-slit spectral study of these galaxies and present now their results.
In the next section we describe our observations, in Section~3 we analyse the galaxy kinematics,
in Section~4 -- the gas excitation and metallicity. In Section~5 we report the stellar population properties
for the central part of the galaxies. In Section~6 we estimate the star formation rates in the rings with the
GALEX UV data. Our results on the characteristics of starforming regions in the ring and possible scenarios
of starforming ring origin are discussed in Section~7.

\section{Observations}
  
The long-slit spectral observations of NGC~1533 and NGC~1543 were made with the Robert Stobie Spectrograph
\citep[RSS;][]{Burgh03,Kobul03} at the 10m Southern African Large Telescope (SALT)
\citep{Buck06,Dono06}, with a slit of 1.25 arcsec width.  The listing of observations is given
in Table~\ref{tbl_logobs}. Since the galaxies look face-on, and we can not determine the orientation
of their lines-of-nodes 'by eye', the slit was posed to catch some blue compact regions in the rings --
possible HII-regions. However for NGC~1543 at $PA=170^{\circ}$ we were unsuccessful to catch outer emission-line
regions. The volume-phase grism PG0900 was used for our program to cover 
the spectral range of 3760$-$6860~\AA\ with a final reciprocal dispersion of
$\approx 0.97$~\AA\ pixel$^{-1}$, except the first spectrum of NGC~1543 which was exposed
in the spectral range of 4190$-$7260~\AA; the spectral resolution was 5.5~\AA.  The
seeing conditions during our observations were poor, in the range of 2.5$-$6.0~arcsec.  The RSS pixel
scale is 0\farcs126, and the effective field of view is 8\arcmin\ along the slit.
We used a binning factor of 4 to get final spatial sampling of 0\farcs502 pixel$^{-1}$.  
Spectrum of an Ar comparison arc was exposed to calibrate the wavelength scale every night
after each galaxy observation as well as spectral flats were observed regularly to correct for
pixel-to-pixel inhomogeneity.  Spectrophotometric standard stars were observed during twilights,
after the observations of the objects, that allowed to correct sensitivity variations along the
dispersion.

\begin{table*}
\centering
\caption{Long-slit spectroscopy of the studied galaxies}
\label{tbl_logobs}
\begin{tabular}{cclccccc}
\hline\noalign{\smallskip}
Galaxy       & Date   &  Exp.      & Binning    &Slit    & PA(slit) & Seeing          \\
	       &           & [sec]      &            &[arcsec]&  [deg]   & [FWHM, arcsec]  \\ \hline\noalign{\smallskip}
NGC\,1533                   & 29.01.2019 & 1300$\times$2      & 2$\times$4 &  1.25   & 50 & 4.6 \\
                            & 18.07.2020 &  1300$\times$2      & 2$\times$4 &  1.25   & 230 & 6.0 \\ \hline
NGC\,1543                  & 02.11.2018 & 1400$\times$2      & 2$\times$4 &  1.25   & 350 & 4.6 \\
                            & 28.12.2018 & 1300$\times$2      & 2$\times$4 &  1.25   & 148 & 2.7 \\ \hline
\end{tabular} 
\end{table*}

Primary data reduction was performed using the standard SALT science pipeline \citep{Cr2010}. The data were then reduced using
the RSS spectral pipeline described in \citet{2022AstB...77..334K}.
The accuracy of the spectral linearisation was checked using the night sky lines [OI]~$\lambda$5577~\AA\  and [OI]~$\lambda$6300~\AA;
the RMS scatter of their wavelengths traced along the slit was 6--8~\kms.  
The sky spectra taken at the slit edges were used to subtract the background.

\section{Kinematical profiles}

The stellar line-of-sight (LOS) velocities and velocity dispersions were calculated by cross-correlating the pixel-to-pixel
galactic spectra with the spectrum of the star HD~58972. Near the galactic centers the profiles have been then
smoothed by the windows corresponding to the seeing during the exposures; farther from the centers the bins
have been augmented to keep sufficient signal-to-noise ratios.

\begin{figure}
	\centering
 	\includegraphics[width=0.45\textwidth]{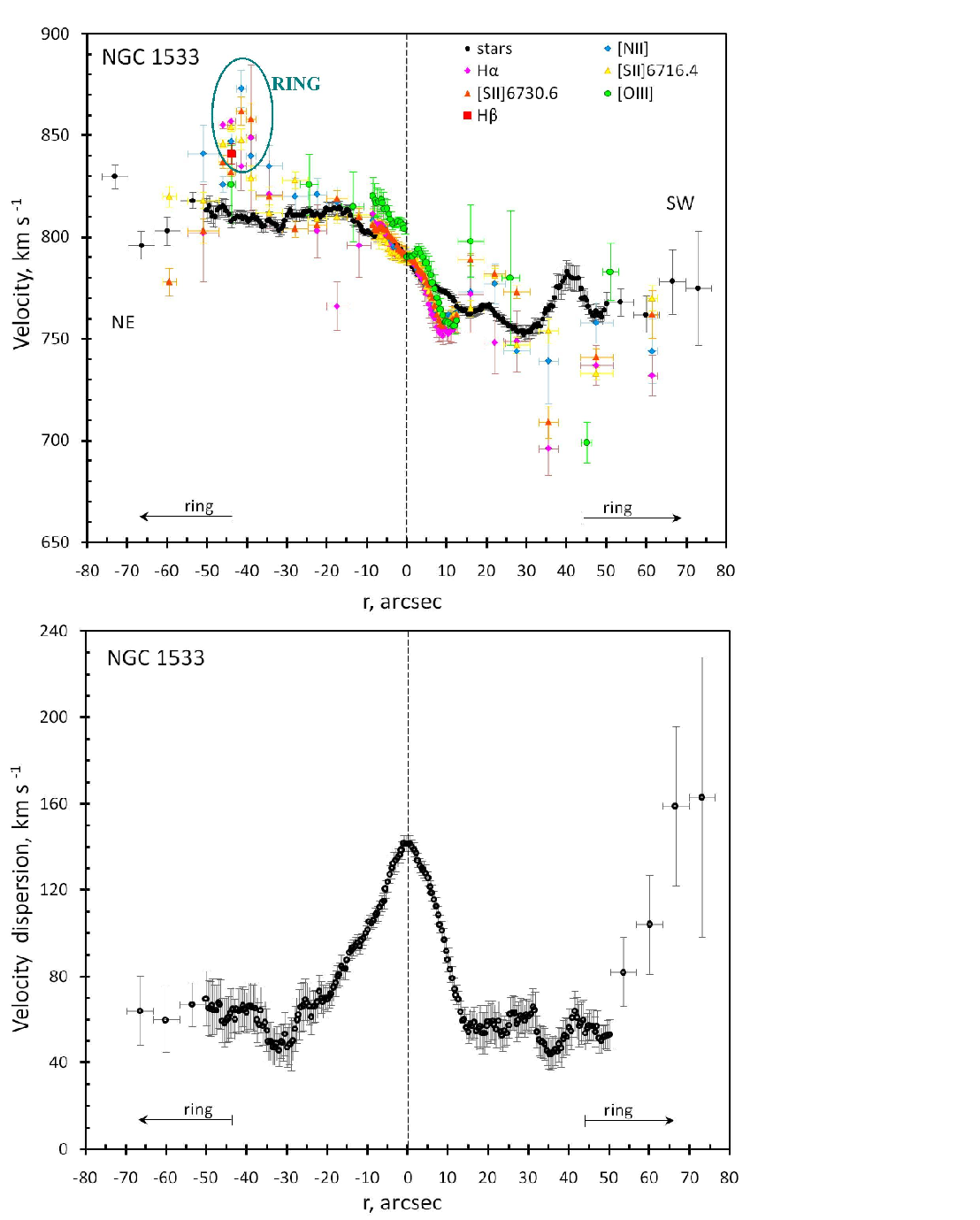}
	\caption{NGC~1533: LOS velocities of stars and ionised gas and the stellar velocity dispersions along the position angle
of $PA=50^{\circ}$.}
	\label{vel1533}
\end{figure}

The cross-section of NGC~1533 along the position angle $PA=50^{\circ}$ chosen rather arbitrarily to probe the north-eastern
HII-region in the ring \citep{SINGG}, shown in Fig.~\ref{vel1533}, reveals the visible rotation of the galaxy that confirms
that NGC~1533 is not exactly face-on that has been implied earlier by the isophote analysis (Table~\ref{tbl_lit}).
The easthern part of the galaxy is approaching in the course of its
rotation, and the stars rotate together with the ionised gas. Strong asymmetry demonstrated both by the ionised-gas
velocity profiles and by the stellar velocity dispersion profile is due perhaps to the bar influence: if the north-eastern
half of the slit catches the bar kinematics it may explain the increased stellar velocity dispersion and shallower
gas LOS velocity gradient because the bar must heat the stellar component and decelerate 
the gas rotation \citep{Roberts1979,Schwarz1985,Combes1990,Athanassoula1992}.
However, at the radius of $R>15^{\prime \prime} -20^{\prime \prime}$ the stellar velocity dispersion reaches the plateau
at the level of some 50--60~\kms demonstrating the domination of the thin stellar disc kinematics. At $R>40^{\prime \prime}$
where the slit crosses the stellar ring, the stellar velocity dispersion rises again.
At the south-west, at $R\approx 43^{\prime \prime}$, the stellar rotation velocity, or more exactly its projection,
drops to zero; this radius is well beyond the bar area but marks the inner edge of the stellar ring. This asymmetric
feature will be attributed by us to possible stellar contribution from outside to be a consequence of minor merger. 
 
\begin{figure*}
	\centering
	\includegraphics[width=0.9\textwidth]{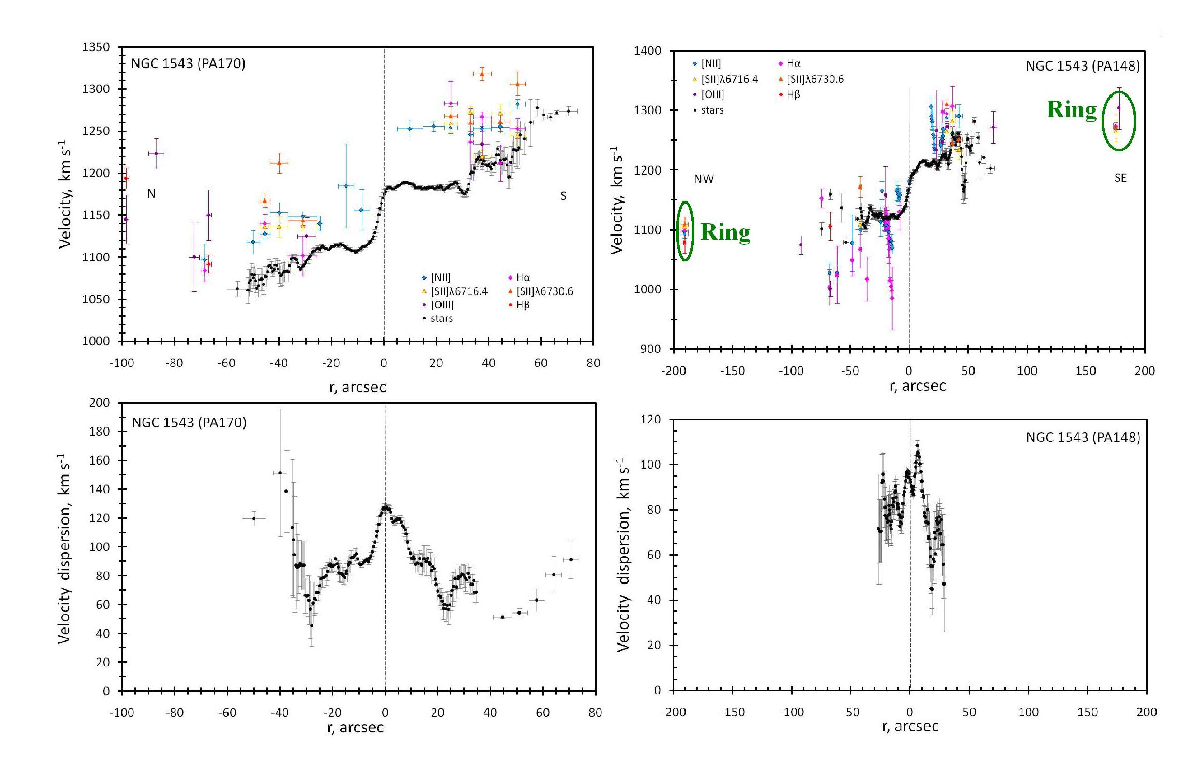}
	\caption{NGC~1543: LOS velocities of stars and ionised gas and the stellar velocity dispersions along the position angle
of $PA=170^{\circ}$ ({\it left}) and $PA=148^{\circ}$({\it right}).}
	\label{vel1543}
\end{figure*}

Among two long-slit cross-sections for NGC~1543, the one along $PA=170^{\circ}$ (Fig.~\ref{vel1543}, left) was taken almost
perpendicularly to the large bar. However, the structure of the galaxy is very complex: besides the large bar
extending in the direction east-west further than one arcminute from the center, the galaxy possesses also the
inner bar at $PA=26^{\circ}$ and with a radius of $11^{\prime \prime}$ \citep{Erwin2004}. It may explain the asymmetry
of both kinematical profiles in Fig.~\ref{vel1543}. Again, similarly to the case of NGC~1533, at $PA=170^{\circ}$
the excess of the stellar velocity dispersion to the south from the center is accompanied by suppressed rotation.
The stellar velocity dispersion remains rather high within the area affected by the bars, but drops to the disc
typical value of 50~\kmps\ at $R\ge 25^{\prime \prime}$ where also the photometric exponential disc starts
to dominate over the dynamically hot galaxy components \citep{Erwin2015}. 
Till the outer part of the disc the stellar velocity dispersion rises again
at $PA=170^{\circ}$ perhaps because of the possible presence of secondary, off-disc stellar component. The peak of
the stellar velocity dispersion shifted to the south-east from the photometric center at $PA=148^{\circ}$
reminds the similar effect in the long-slit cross-section of NGC~1543 taken along the large bar, at $PA=270^{\circ}$,
by \citet{Jarvis1988}.

If we look at the flat parts of the velocity profiles at $R\ge 60^{\prime \prime}$ where non-circular motions due to
the bars may be already consider negligible, we fix the amplitude of the projected rotation velocity of 100~\kms\ at
$PA=170^{\circ}$ while at $PA=148^{\circ}$ it is only $\sim 65$~\kms. Hence we must conclude that $PA=170^{\circ}$
is closer to the line of nodes of the stellar disc than $PA=148^{\circ}$. This conclusion is consistent with
the orientation of the photometric major axis of the outer isophotes found in the Carnegie-Irvine photometric 
survey (Table~\ref{tbl_lit}) but disagrees with the orientation of the neutral-hydrogen disc rotation plane implying
the kinematical major axis -- the orientation of the line of nodes of the large-scale gaseous disc (outer ring) --
at the $PA=134^{\circ}$ \citep{Murugeshan2019}.
Meantime the ionised-gas velocities within the radius of $R\sim 80^{\prime \prime}$ in Fig.~\ref{vel1543} do 
not diverge with the stellar velocities though the emission lines are there very weak and difficult to be measured
exactly.

\section{Ionised gas in the rings of NGC~1533 and NGC~1543}

\subsection{Excitation}

\begin{figure*}
\begin{tabular}{c c}
 \includegraphics[width=0.45\textwidth]{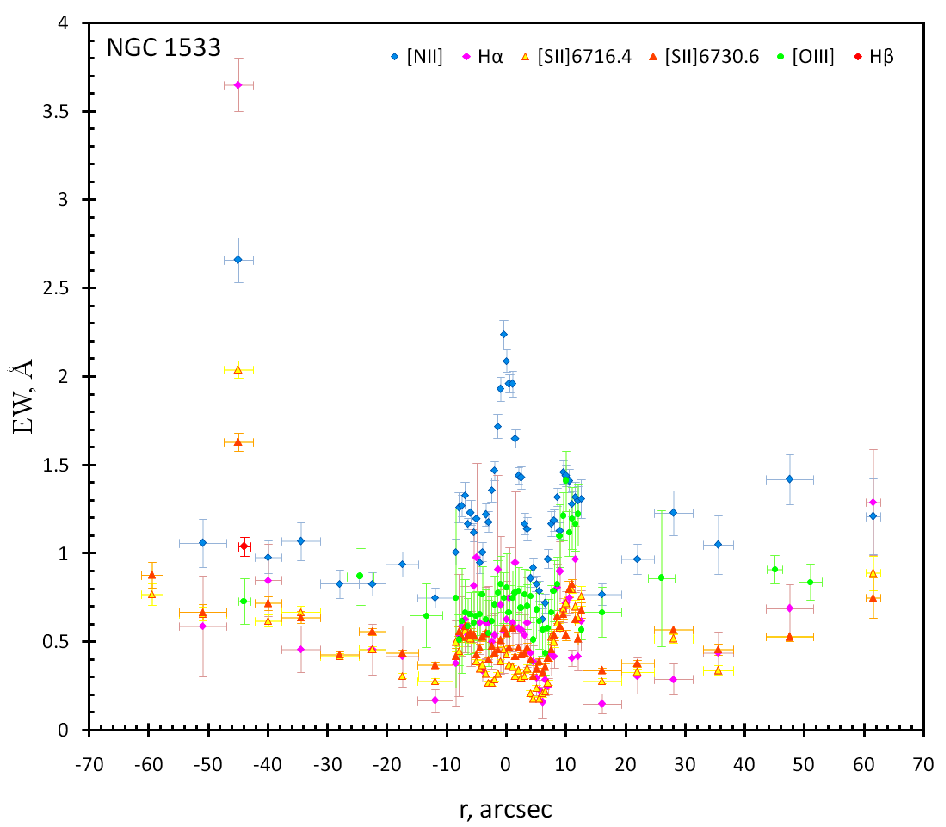} &
 \includegraphics[width=0.45\textwidth]{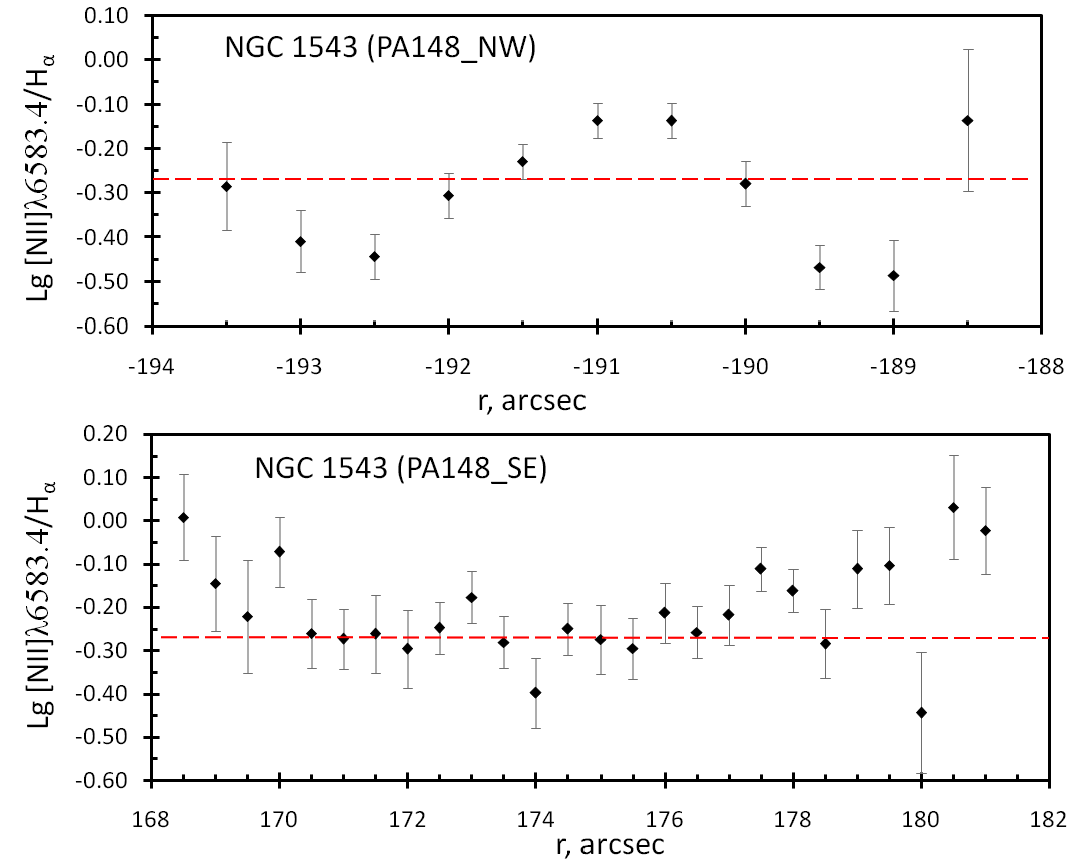} \\
\end{tabular}
\caption{The profiles: of the emission-line equivalent widths in NGC~1533 along the cross-section
in $PA=50^{\circ}$ ({\it the left plot}) and of the nitrogen-to-hydrogen line flux ratio in NGC~1543 
in the 3-arcmin ring in $PA=148^{\circ}$ ({\it the right plot}). 
}\label{lineprofs}
\end{figure*}

Firstly we have calculated fluxes and equivalent widths of the emission lines in our spectra along the radii in NGC~1533 and NGC~1543;
we present our measurements in some selected radius ranges in Fig.~\ref{lineprofs}.

\begin{figure*}
\begin{tabular}{c c}
 \includegraphics[width=0.45\textwidth]{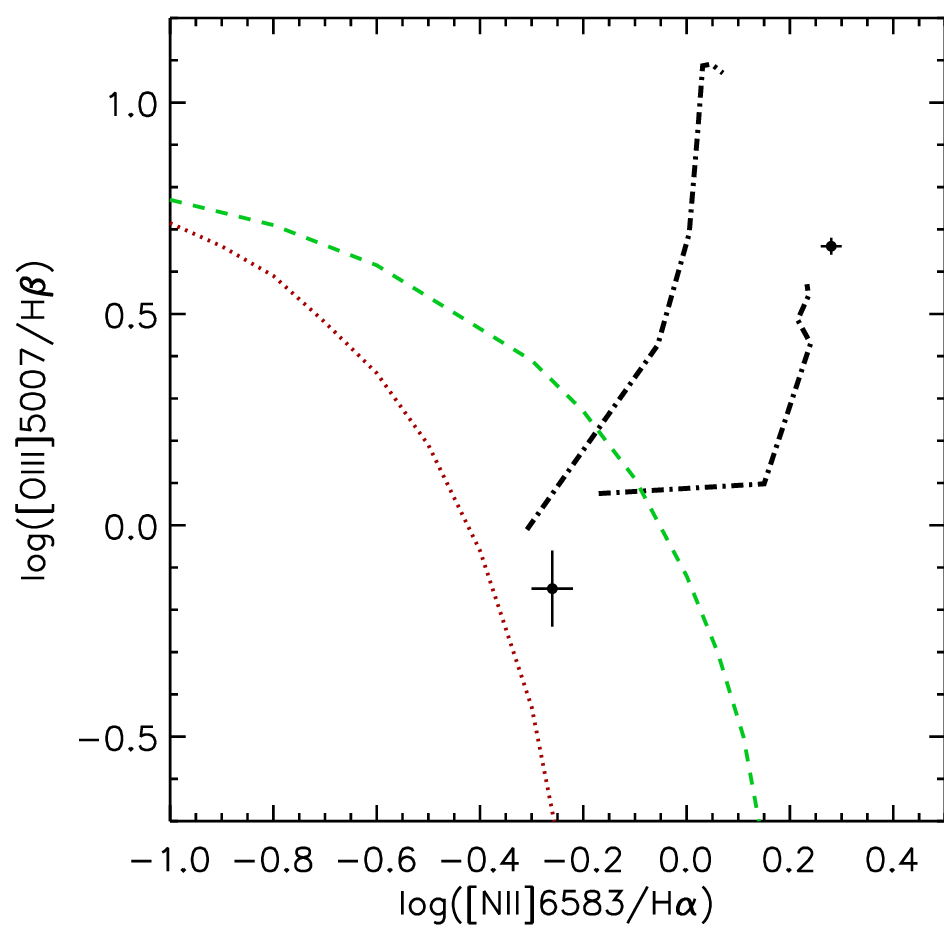} &
 \includegraphics[width=0.45\textwidth]{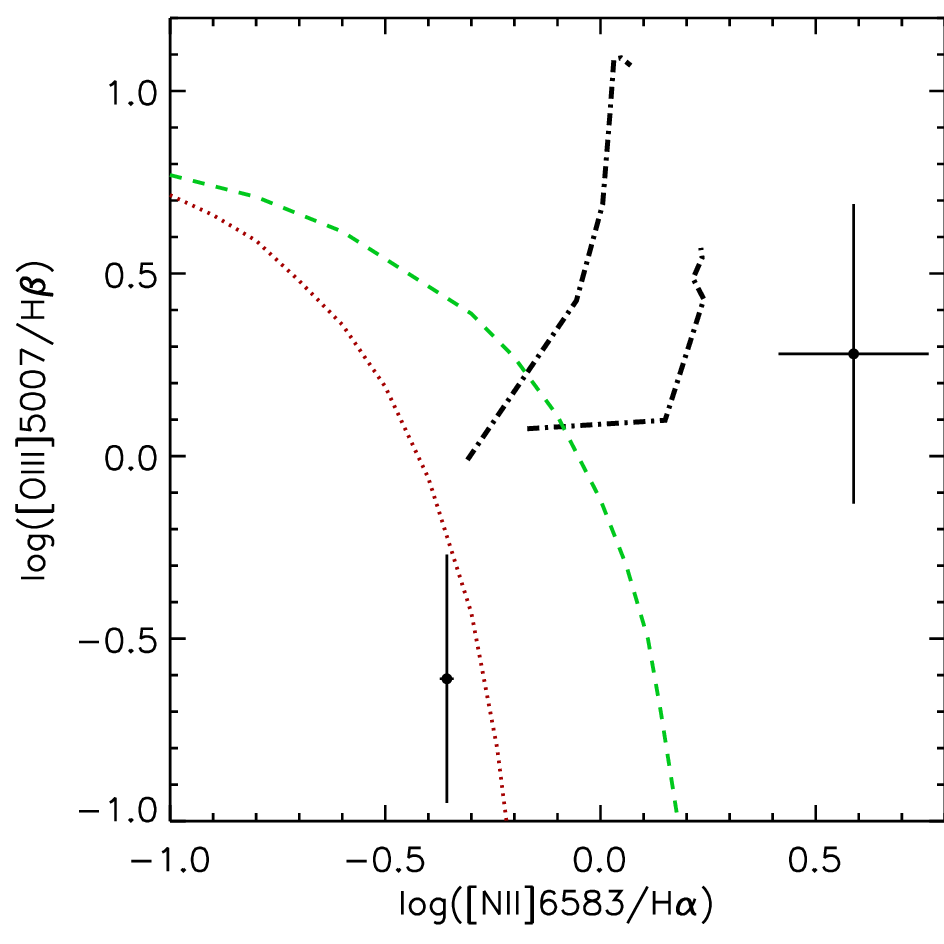} \\
\end{tabular}
\caption{
Baldwin-Phillips-Terlevich (BPT) diagrams for the ionised gas in the rings. Excitation-mechanism 
dividing lines are from \citet{kewley01} (the green dashed line) and from \citet{kauffman_2003} (the red dotted line), 
and the dashed-dot lines show the shock excitation models by \citet{allen_shock}, with the shock wave velocity varying
from 100~\kms\ to 1000~\kms\ from bottom to top.
{\it The left plot} -- NGC~1533, the nucleus is in the LINER area of the BPT-diagram and the north-eastern HII-region in the ring
is in the composite area where excitation by young stars and shock waves may work together; though both can be explained by shock excitation
of the gas without precursor while assuming the shock velocity of 100~\kms\ for the ring and 1000~\kms\ for the nucleus.
{\it The right plot} -- NGC~1543, $PA=148^{\circ}$, with the emission-line flux ratios measured in two emission-line regions,
the inner one ($R=43^{\prime \prime}$) being in the shock-dominated area and the outer one (at the ring radius of $R=193^{\prime \prime}$)
being in the HII-region-type excitation area.
}\label{bpt_gal}
\end{figure*}

The emission lines in NGC~1533 are strong in the nucleus and in the northern arc of the ring, at $R=45^{\prime \prime}$
from the center. The northern emission-line region in the ring of NGC~1533 was earlier designated as the B-region 
by \citet{Rampazzo2020} who detected it in the narrow-band photometric data. The radial variations of the line equivalent
widths are shown in Fig.~\ref{lineprofs}, {\it left}, and one can see that the strongest emission line in the nucleus 
is [NII]$\lambda$6583 and in the ring it is the hydrogen H$\alpha$. We have inspected the ionised-gas excitation
mechanisms with so called Baldwin-Phillips-Terlevich (BPT) diagram \citep{BPT} in Fig.~\ref{bpt_gal}({\it left}) and have assured that
the nucleus of NGC~1533 is rather a weak LINER while the B-region is in the 'composite' area of the BPT-diagram: it
may be excited both by young stars and shock waves together. Meantime both can be also explained by pure shock excitation
of the gas without precursor while assuming the shock velocity of 100~\kms\ for the ring and 1000~\kms\ for the nucleus.
Beyond the nucleus and the ring the equivalent widths
of the lines including H$\alpha$ are mostly below 1~\AA\ so they can be excited by old stars \citep{oldstar_excitation}.

The emission lines in the red spectral range of the NGC~1543 spectra are weak everywhere except two compact locations
at the slit in $PA=148^{\circ}$, with the enhancements at $R\approx 43^{\prime \prime}$ and another one at $R=3.2^{\prime}$,
within the outer ring; hence the radius of the outer emission-line ring reaches 20~kpc.
The emission lines in the green spectral range are even weaker, and though the emission lines are seen in both
northern and southern segments of the $PA=148^{\circ}$ cross-section, we have been lucky to measure [OIII]$\lambda$5007
only to the north from the center. At the diagnostic BPT-diagram (Fig.~\ref{bpt_gal}, {\it right}), one can see that
the inner emission-line region is excited by a very fast shock wave, and the outer ring contains ionised gas excited by young stars.
By fixing the ratio $\lg (\mbox{[OIII]}\lambda 5007 / \mbox{H}\beta )=-0.6$
-- at the the value which has been measured in the north-western segment of the outer ring, --
we can put a restriction onto the ratio $N2\equiv \lg (\mbox{[NII]}\lambda 6583 / \mbox{H}\alpha ) <-0.27$ for the
gas excited by young stars through using the condition formulated by \citet{kewley06}. Then in Fig.~\ref{lineprofs}, {\it right},
we can recognise that the south-eastern segment of the outer ring is excited by young stars in the radius range of
$170^{\prime \prime} - 177^{\prime \prime}$.

\subsection{Oxygen abundance in the ionised gas}

When the gas is excited by young stars we can determine its metallicity by using the model calibrations of
the line flux ratios under assumptions of different oxygen abundances. We use here two indicators which we can measure:
$N2\equiv \lg (\mbox{[NII]}\lambda 6583 / \mbox{H}\alpha )$ and 
$O3N2\equiv  \lg (\mbox{[OIII]}\lambda 5007 / \mbox{H}\beta ) - \lg (\mbox{[NII]}\lambda 6583 / \mbox{H}\alpha )$
calibrated by \citet{pettini_pagel} and later refined by \citet{marino}. Since in the ring of NGC~1533
the gas excitation mechanism is composite for this galaxy we use only the indicator $O3N2$ which works well even
for the composite BPT-diagram area as has been shown by \citet{Kumari2019}. Then in NGC~1533 the oxygen abundance
of the gas in the emission-line ring is $12+\log \mbox{O/H} = 8.70 \pm 0.04$ according to \citet{pettini_pagel} 
and $8.51\pm 0.04$ according to \citet{marino}. For NGC~1543 the indicator $N2$ implies the oxygen abundance
in the north-western segment of the outer ring to be $12+\log \mbox{O/H} = 8.68 \pm 0.04$ according to \citet{pettini_pagel}
calibration and $12+\log \mbox{O/H} = 8.57 \pm 0.03$ according to calibration by \citet{marino}, and for
the south-eastern tip of the outer ring the corresponding estimates are $8.73\pm 0.01$ and $8.60\pm 0.01$.
Here we give the statistical accuracy related to the accuracy of the line flux ratio determination; the accuracy
(rms scatter) of the calibrations are larger, about 0.18~dex. Taking in mind that the solar oxygen abundance
is 8.69 \citep{Asplund2009}, we conclude that the ionised gas in the outer rings of NGC~1533 and NGC~1543 has
solar metallicity or slightly subsolar one -- in agreement with our previous results for a sample of outer starforming
rings in lenticular galaxies \citep{s0_fp,Proshina2019}.

\subsection{Background galaxies projected onto the rings}

\begin{figure*}
\begin{tabular}{c c}
 \includegraphics[width=0.65\textwidth]{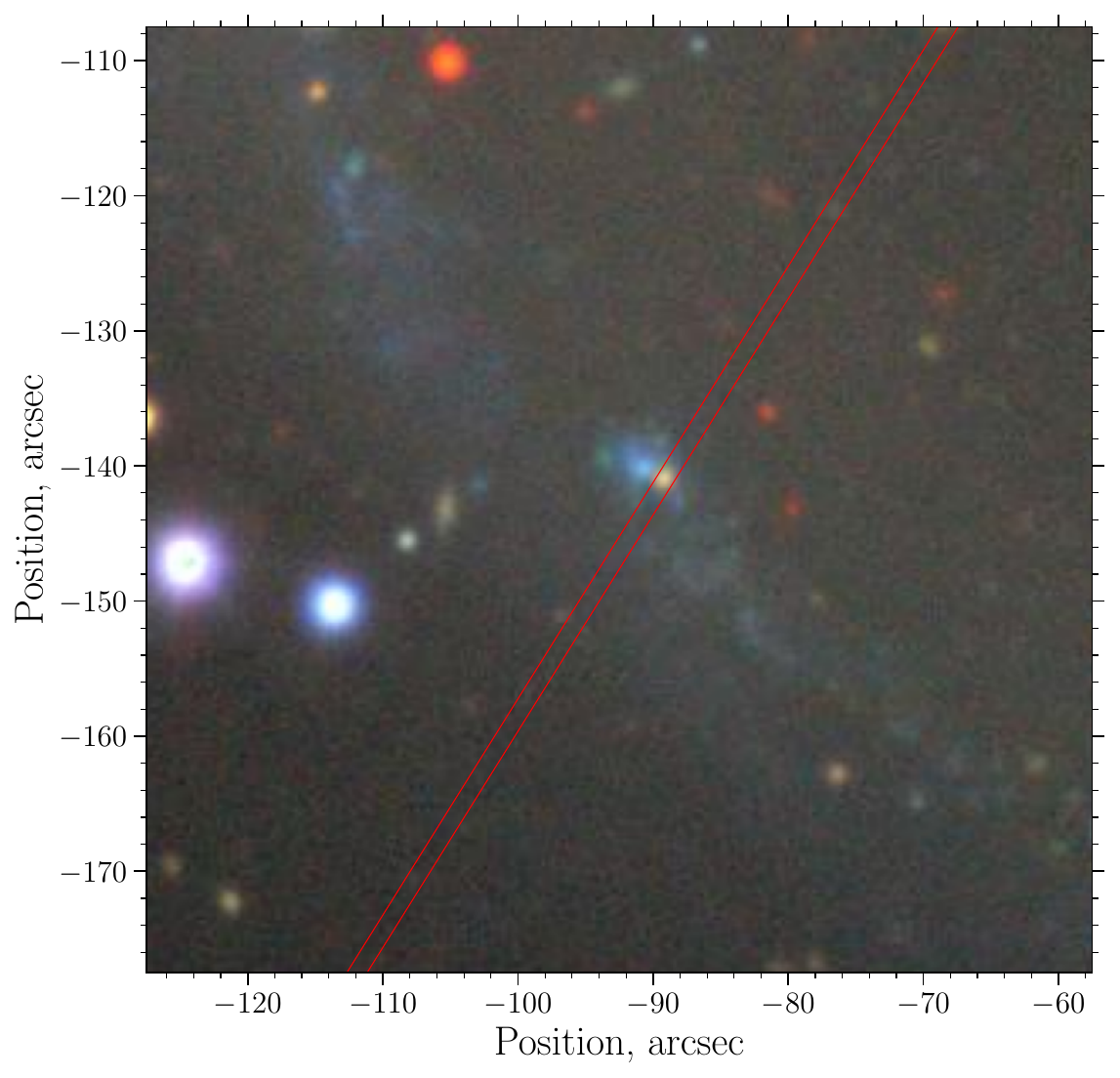} &
 \includegraphics[width=0.35\textwidth]{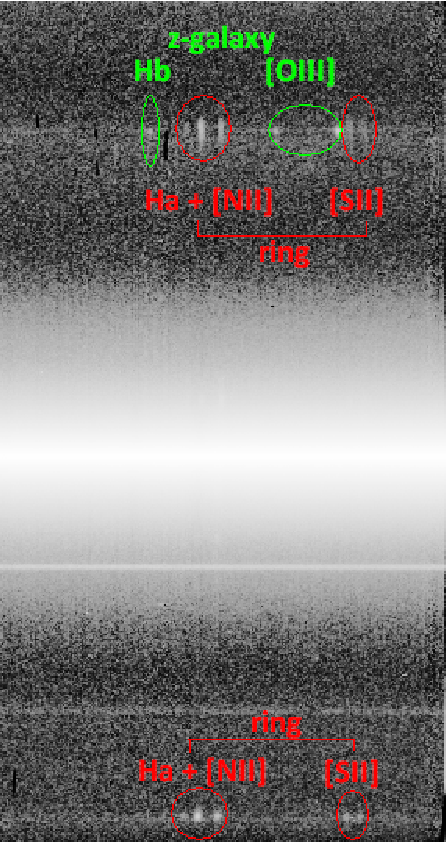} \\
\end{tabular}
\caption{The background galaxy projected just against the outer ring of NGC~1543: {\it the left plot} -- a south-eastern
part of the NGC~1543 image from the DECaLS survey with the background galaxy just in the slit,
{\it right plot} -- the direct view of the red part of the spectrum of NGC~1543 where at the top the background galaxy is projected onto
the ring of NGC~1543 with its emission line H$\beta$ to the left from the ring H$\alpha$ and its [OIII]$\lambda$4959,5007 
doublet to the right from the ring [NII]$\lambda$6583.
}\label{fongal}
\end{figure*}

While inspecting the emission lines in the ring of NGC~1543 we have found a curious overlapping: emission lines
of background galaxy have projected through the south-eastern part of the outer ring of NGC~1543 and have been
catched by our long-slit cross-section at $PA=148^{\circ}$. Figure~\ref{fongal} shows the direct coloured image of
this galaxy at the left plot. It is a background compact starforming galaxy (Fig.~\ref{fongal}, {\it left}) just in the slit.
In the spectrum, at the distance of $3^{\prime}$ from the center of NGC~1543 we see compact emission-line clumps including
the lines H$\alpha$ and [NII]$\lambda$6548,6583 belonging to the ring of NGC~1543 and the lines [OIII]$\lambda$4959,5007 and H$\beta$
belonging to the background galaxy (Fig.~\ref{fongal}, {\it right}). The latter three emission lines have allowed us
to measure the redshift of this background galaxy WISEA J041254.97-574643.6: it is equal to $z=0.3447\pm 0.0001$.

\begin{figure*}
\begin{tabular}{c c}
 \includegraphics[width=0.45\textwidth]{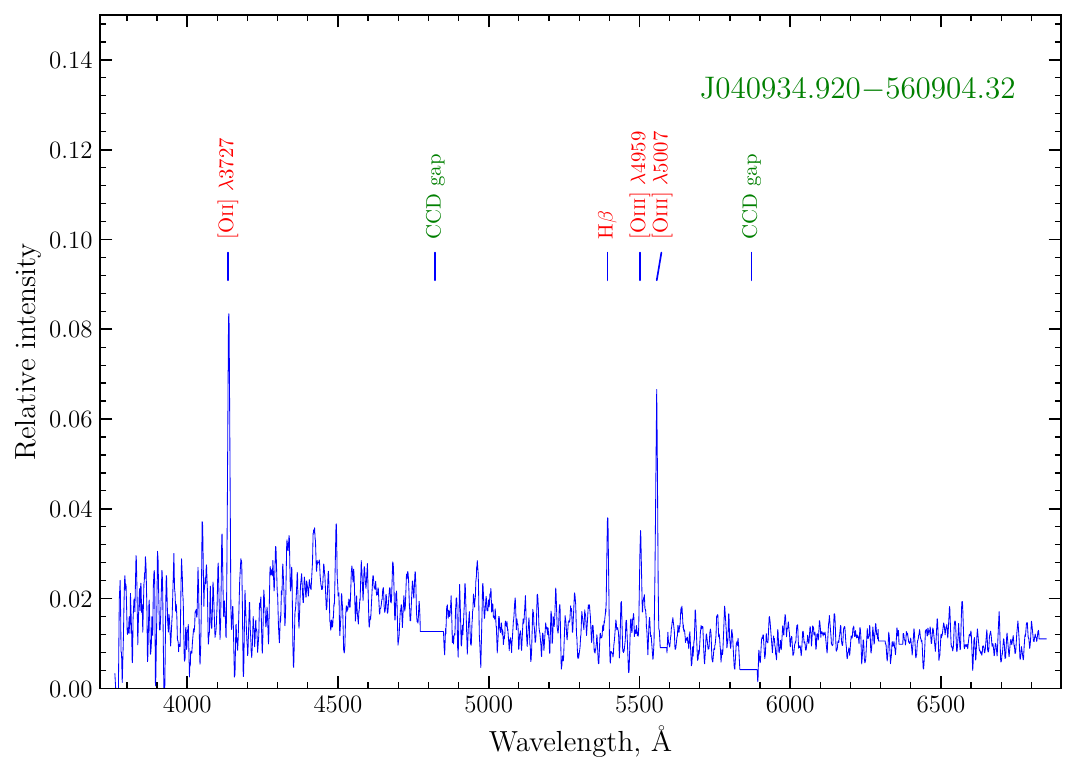} &
 \includegraphics[width=0.45\textwidth]{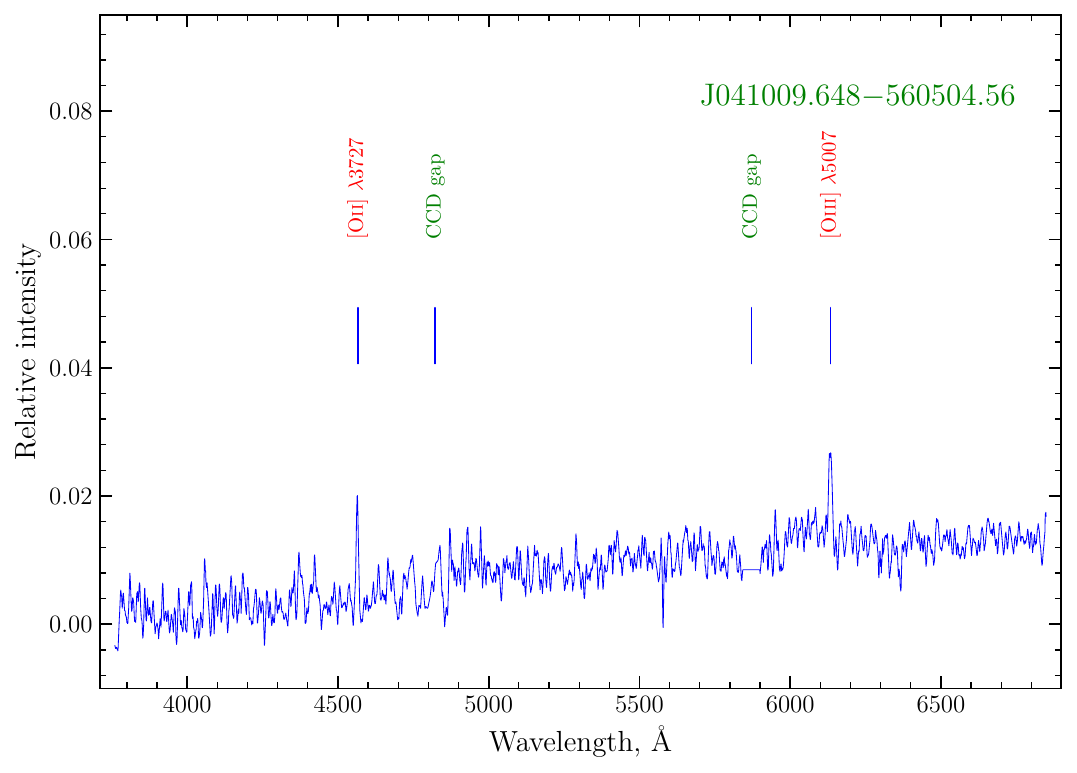} \\
\end{tabular}
\caption{The integrated spectra of two background galaxies catched by the slit during the long-slit observations of NGC~1533.
The strong emission lines are marked as well as the pieces of the spectra where the signal was missed due to the gap in the CCD
mosaic.}
\label{sp_dwarfs}
\end{figure*}

Besides this rather bright and emission-rich background galaxy at the slit through NGC~1543, we have noticed several
faint background dwarf galaxies projected against the spectral cross-section through NGC~1533, also with some emission lines
providing the possibility to measure their spectral redshifts. The integrated spectra of two of them are shown in Fig.~\ref{sp_dwarfs},
and some photometric and spectral characteristics are presented in Table~\ref{tbl:addgal}. As it can be seen in Table~\ref{tbl:addgal}
the photometric redshift of WISEA J041254.97-574643.6 --  of the brightest galaxy, -- which has been derived according to the prescriptions
by \citet{duncan2022}, disagrees strongly, well beyond the formal accuracy, with the spectroscopic measurement.
Evidently, this discrepancy is due to the clumpy foreground provided by NGC~1533 and NGC~1543.

\begin{table*}
	\caption{Background galaxies found in RSS long-slit spectra used in this paper}
	\label{tbl:addgal}
	\begin{tabular}{ccccl} \hline
		SALT Name               &  $r$ mag$^a$ & Phot.\,redshift$^a$ &  SALT redshift    &  Identified Lines                                         \\ \hline
		J040934.920$-$560904.32 &  22.07       & 0.256$\pm$0.196     &  0.1039$\pm$0.0002& [\ion{O}{ii}]~$\lambda$3727, H$\beta$, [\ion{O}{iii}]~$\lambda$4959,5007 \\
		J041001.752$-$560557.84 &  23.24       & ---                 &  0.3411$\pm$0.0010& [\ion{O}{ii}]~$\lambda$3727                               \\
		J041009.648$-$560504.56 &  19.11       & 0.228$\pm$0.028     &  0.1998$\pm$0.0003& [\ion{O}{ii}]~$\lambda$3727, H$\beta$, [\ion{O}{iii}]~$\lambda$5007 \\
		J041254.970$-$574643.60 &  17.37       & 0.045$\pm$0.032     &  0.3447$\pm$0.0001& [\ion{O}{ii}]~$\lambda$3727, H$\beta$, [\ion{O}{iii}]~$\lambda$4959,5007 \\
		\hline
		\MC{5}{p{12.0cm}}{$^a$ -- $r$ magnitudes and photometric redshifts are taken from Legacy Survey data \citep{legacysurvey}}
	\end{tabular}
\end{table*}

\section{Stellar population properties}

\begin{figure*}
	\centering
	\includegraphics[width=0.9\textwidth]{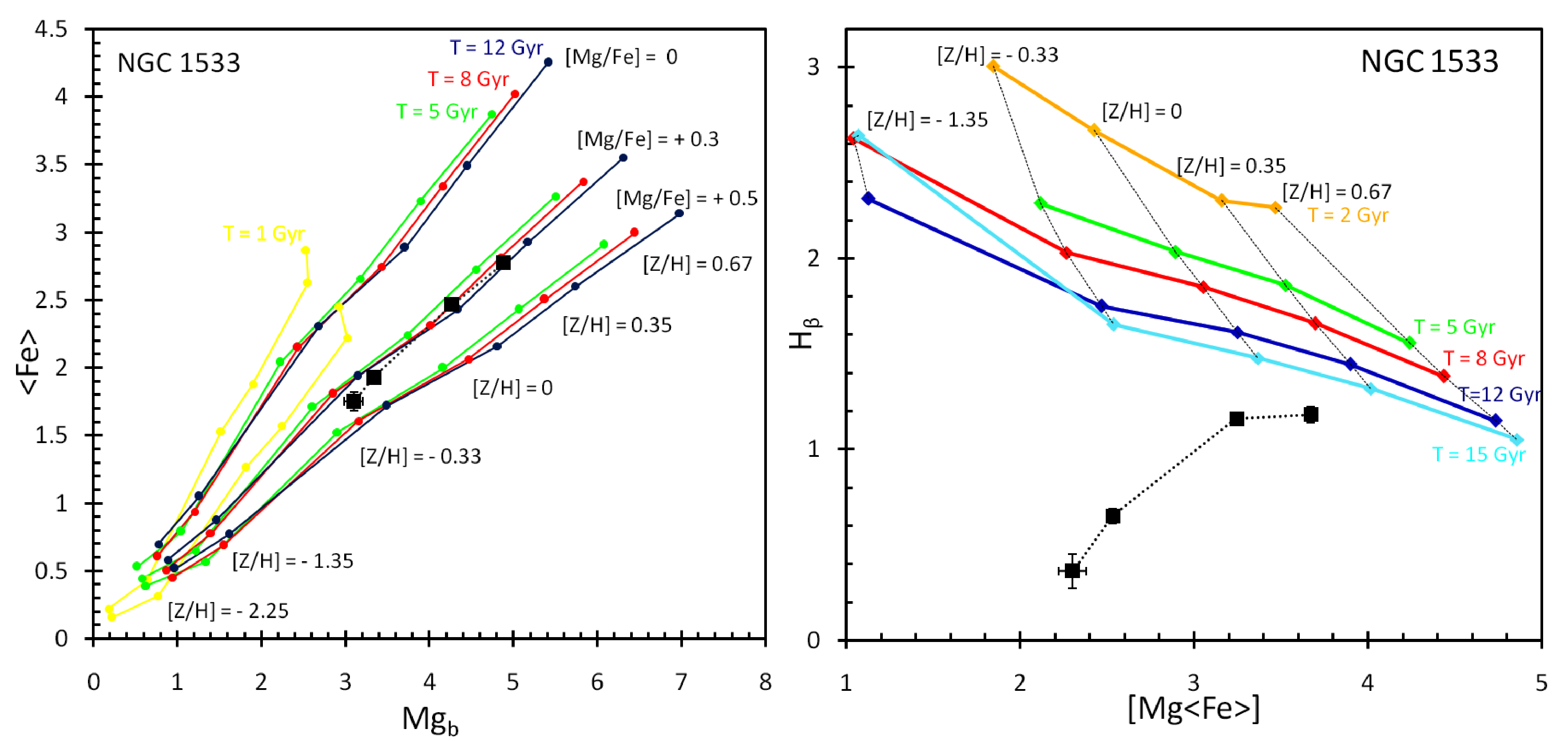}
	\caption{Lick index-index diagrams for NGC~1533.
The {\it left plot} represents Mgb vs iron index diagram which allows to estimate magnesium-to-iron ratio through the comparison
of our measurements with the models by \citet{thomod} for the different Mg/Fe ratios. By confronting the H$\beta$ Lick index versus
a combined metallicity Lick index involving magnesium and iron lines ({\it right plot}), we solve the metallicity-age degeneracy
and determine these stellar-population parameters with the SSP evolutionary synthesis models by \citet{thomod}.
Five different age sequences are plotted by coloured curves as a reference frame; the dashed lines crossing the model age
sequences mark the metallicities of $+0.67$, $+0.35$, 0.00, --0.33 from right to left. The black squares represent
our measurements for NGC~1533 resolved along the radius. From the right end of the broken dotted sequence
connecting these black squares, by fixing large radial bins,
we go along the radius through the galaxy structure components: the nucleus is taken at $R=0^{\prime \prime}-3^{\prime \prime}$,
the bulge and the bar are taken at $R=4^{\prime \prime}-10.5^{\prime \prime}$,
the inner disc inside the ring is taken at $R=11^{\prime \prime}-40^{\prime \prime} $, and the outer ring is fixed at $R >40^{\prime \prime}$.
}
\label{lick1533}
\end{figure*}

\begin{figure*}
	\centering
	\includegraphics[width=0.9\textwidth]{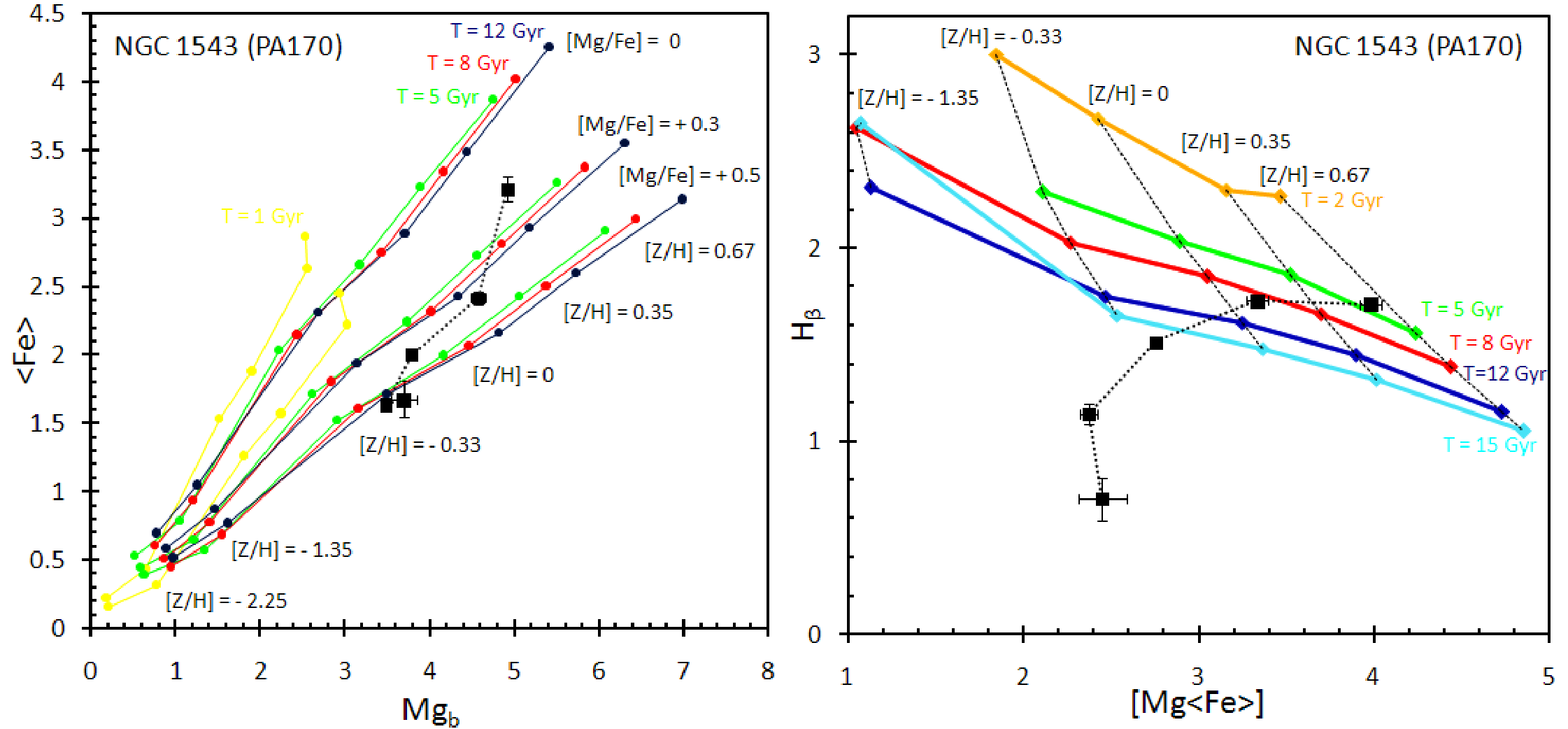}
	\caption{Lick index-index diagrams for NGC~1543.
The {\it left plot} represents Mgb vs iron index diagram which allows to estimate magnesium-to-iron ratio through the comparison
of our measurements with the models by \citet{thomod} for the different Mg/Fe ratios. By confronting the H$\beta$ Lick index versus
a combined metallicity Lick index involving magnesium and iron lines ({\it right plot}), we solve the metallicity-age degeneracy
and determine these stellar-population parameters with the SSP evolutionary synthesis models by \citet{thomod}.
Five different age sequences are plotted by coloured curves as a reference frame; the dashed lines crossing the model age
sequences mark the metallicities of $+0.67$, $+0.35$, 0.00, --0.33 from right to left.
The black squares represent our measurements for NGC~1543 resolved along the radius. From the right end of the broken
dotted sequence connecting these black squares, by fixing large
radial bins, we go along the radius through the galaxy structure components: the nucleus is restricted to
$R=0^{\prime \prime}-1.5^{\prime \prime}$, the central core is taken at $R<4^{\prime \prime}$, and then we show the bulges measured within
$R=4.5^{\prime \prime}-16^{\prime \prime}$, the inner disc inside the first (inner) emission-line ring taken at
$R=16.5^{\prime \prime}-31.5^{\prime \prime}$, and the part of the inner disc outside the first (inner) emission-line ring taken in the
range of $R=32^{\prime \prime}-70^{\prime \prime}$.
}
\label{lick1543}
\end{figure*}

NGC~1533 and NGC~1543 are both barred lenticular galaxies with UV-bright, star-forming outer rings.
The source of gas for star formation and the trigger of its start are topics for our investigation.
Have this gas come from outside, or is it primordial gas of these galaxies re-distributed by the bar
toward the outer Lindblad resonances? The star formation histories of the inner parts of the galaxies
must be connected with the answers to these questions. 

We have studied the stellar population parameters of the inner parts of NGC~1533 and NGC~1543 by calculating
Lick indices \citep{Worthey1994} H$\beta$, Mgb, Fe5270, and Fe5335 along the slit. The index H$\beta$ was corrected
for the emission-line contamination by using the relation $EW(\mbox{H}\beta)=0.25 EW(\mbox{H}\alpha)$ found
by \citet{SS2001} for a large inhomogeneous sample of spiral galaxies -- since the ionised gas
in the central parts of NGC~1533 and NGC~1543 is everywhere excited by shocks, or by old stars, and Balmer
decrements calculated for HII-regions, so called B-model for hydrogen recombination \citep{Burgess1958}, is not
relevant in this case. The instrumental SALT/RSS Lick indices are calibrated into the standard Lick index
system \citep{woretal} with the relations established by \citet{katkov_salt} for the same spectrograph configuration.

The Lick indices calculated over the wide radial bins corresponding to the nuclei, bulges, bar-dominated
regions, and to the inner and outer discs, are plotted in Fig.~\ref{lick1533} and in Fig.~\ref{lick1543}.
We have identified the radial ranges dominated by one or another structure component by applying
photometric decomposition to the galaxies images obtained in the near-infrared by the space telescope Spitzer and retrieved from
the S4G survey \citep{sheth10} data archive. The detailed analysis of the galaxies structures will be given below, in the Discussion section.
We compare our Lick-index measurements to the Simple Stellar Population (SSP) models calculated by \citet{thomod} for varying
characteristics of stellar populations: ages, metallicities, magnesium-to-iron ratios. The plots
confronting Mgb to the combined iron index, $\langle \mbox{Fe} \rangle \equiv (\mbox{Fe5270} + \mbox{Fe5335})/2$,
allow to estimate the magnesium-to-iron ratios which are the indicators of the star-formation epoch duration \citep{chem86}:
the magnesium overabundance with respect to the solar Mg/Fe value implies the duration of the star formation epoch less than
1~Gyr, finishing before the SNeIa contribution to the iron nucleosynthesis in the galaxy occurs to be significant.
One can see that in NGC~1533 (Fig.~\ref{lick1533}, {\it left}) the Mg/Fe ratio is twice the solar one over 
the whole extension of the galaxy: star formation had completely finished after maximum 1~Gyr from the beginning, so this galaxy
was never a spiral one, with extending star formation history in the disc.
The nucleus, the bulge, and the disc in NGC~1533 are all older than 12~Gyr (Fig.~\ref{lick1533},
{\it right}). In NGC~1543 the magnesium-to-iron ratio is not so large in its nucleus (Fig.~\ref{lick1543}, {\it left}),
and the SSP-equivalent age of the nucleus is less than 5~Gyr (Fig.~\ref{lick1543}, {\it right}). According to the calibration
by \citet{rsmith09}, the SSP-age of 5~Gyr may correspond to the star formation quenching only 2~Gyr ago, so this galaxy
has probably experienced the nuclear re-juvenation very recently. However the large-scale structure components of NGC~1543,
the bulge and the disc, are old in this galaxy too.

\section{Star formation in the rings}

Stars dominating in the ultraviolet radiation of stellar populations are of O- and B-type, so being rather massive, they
indicate recent star formation. By treating the FUV ($\lambda=154$~nm) and NUV($\lambda=232$~nm) fluxes measured for the
galaxies by the space telescope GALEX, \citet{kennrev} have calibrated star formation rates averaged over
the last 100 and 200 Myr, respectively.

\begin{figure*}
	\centering
	\includegraphics[width=0.9\textwidth]{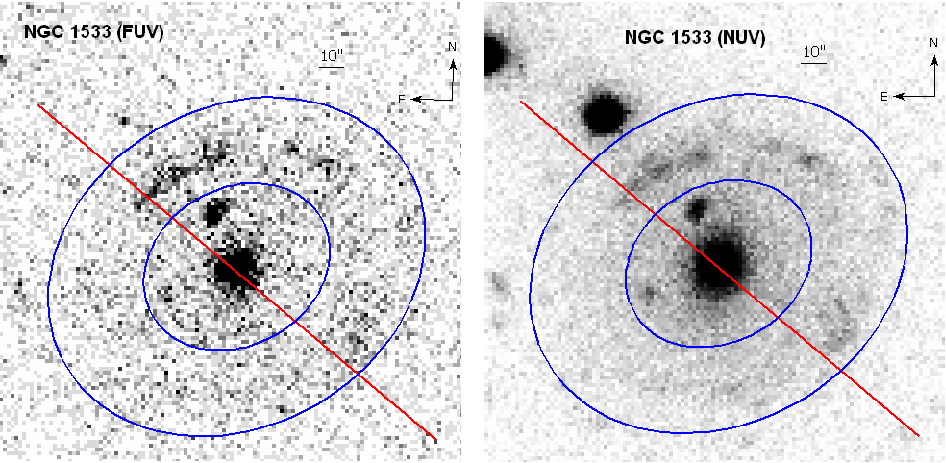}
	\caption{The GALEX maps of NGC~1533 in the FUV- and NUV-bands. The ring aperture is overplotted
where we have calculated the star formation rate.
}
\label{uv1533}
\end{figure*}

\begin{figure*}
	\centering
	\includegraphics[width=0.9\textwidth]{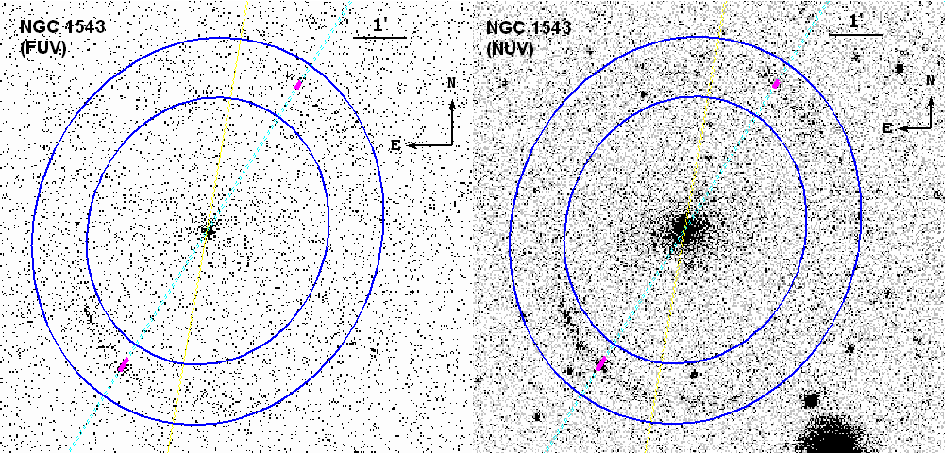}
	\caption{The GALEX maps of NGC~1543 in the FUV- and NUV-bands. The ring aperture is overplotted
where we have calculated the star formation rate.
}
\label{uv1543}
\end{figure*}

We have retrieved the GALEX data for NGC~1533 and NGC~1543 in the MAST archive. The log of the GALEX observations for
these galaxies is given in Table~\ref{tbl_loguv}. Both galaxies reveal UV-bright rings matching the emission-line
loci at the radius of $\sim 50^{\prime \prime}$ in NGC~1533 and of $\sim 3^{\prime}$ in NGC~1543 (Fig.~\ref{uv1533} and
Fig.~\ref{uv1543}). The GALEX data for NGC~1533 and integrated FUV- and
NUV-magnitudes for this galaxy were earlier reported by \citet{marino11}.

\begin{table*}
\centering
\caption{Ultraviolet GALEX observations for the studied galaxies}
\label{tbl_loguv}
\begin{tabular}{ccccr}
\hline\noalign{\smallskip}
Galaxy & Band & Program identifier & Date   &  Exposure time, sec \\ \hline\noalign{\smallskip}
NGC 1533 & FUV & AIS-420-0001-sg06 & 08.10.2007 &  96 \\
\cline{2-5}
                             & FUV & AIS-425-0002-sg93 & 17.11.2007 &  157 \\
                             \cline{2-5}
                             & FUV & GI3-087004-NGC1533-0003 & 02.01.2007 &  1517 \\
                             \cline{2-5}
                             & NUV & AIS-420-0001-sg06 & 08.10.2007 &  192 \\
                             \cline{2-5}
                             & NUV & AIS-425-0002-sg93 & 17.11.2007 &  233 \\
                             \cline{2-5}
                             & NUV & GI3-087004-NGC1533-0001 & 18.11.2007 &  3146 \\ \hline

NGC 1543 & FUV & AIS-420-0001-sg12 & 07.10.2007 &  168  \\
\cline{2-5}
                             & NUV & AIS-420-0001-sg12 & 07.10.2007 &  168 \\ \hline
\end{tabular} 
\end{table*}

We have selected with elliptical apertures the rings at the FUV- and NUV-maps and have calculated the summed ring fluxes in the FUV
and in the NUV bands through these apertures for every galaxy by using the software SAOImage DS9. For NGC~1543, stars projected 
onto the ring area were carefully masked. The initial counts were re-calculated into AB-magnitudes by following the prescriptions 
of \citet{Morrissey_GALEX}. Those were corrected for the extinction in our Galaxy by firstly taking the $A_B$ from \citet{dust2011}, 
and then transforming the $A_B$ into $A_{FUV}$ and $A_{NUV}$ through the formulae from \citet{lee09,lee11}.
The intrinsic absorption within the galaxies was taken into account by involving the infrared images of the galaxies provided by the
WISE survey in the W4-band (22$\mu$m).

Then we calculated the UV-luminosities by taking into account the distances to the galaxies listed in Table~\ref{tbl_lit}. 
The star formation rates (SFR) were finally calculated by using the calibrations of \citet{kennrev}.
The results are compiled in Table~\ref{tbl_uvres}. We would note that the star formation rate integrated
over the whole galaxy NGC~1533 was earlier reported by \citet{Rampazzo2021} with the FUV data derived from the observations
of the space telescope Astrosat/UVIT. If we take into account our accepted distance to NGC~1533, 20.89~Mpc, different from one
assumed by \citet{Rampazzo2021}, 17.69~Mpc, we would come to the UVIT estimate $SFR=0.053 \pm 0.008$\Msunyr, which is consistent
with our estimate. 

\begin{table*}
\centering
\caption{Star formation rates in the rings determined from the GALEX data}
\label{tbl_uvres}
\begin{tabular}{cccc}
\hline\noalign{\smallskip}
Galaxy & Ring borders, $^{\prime \prime}$ &  SFR,\Msunyr, in FUV & SFR,\Msunyr, in NUV \\ \hline\noalign{\smallskip}
NGC 1533 & 38--77 & $0.0386\pm 0.0006$ & $0.04966\pm 0.00002$ \\
NGC 1543 & 149--216 & $0.0823\pm 0.0015$ & $0.08513\pm 0.00071$ \\ \hline
\end{tabular} 
\end{table*}

Interestingly, the SFR estimates made basing separately on the FUV- and NUV-luminosities for the rings are the same in NGC~1543,
but differ by a factor of 1.3 (29\%) in NGC~1533, though the UV-data for NGC~1533 have a better accuracy because of the longer
exposures. Far-UV and near-UV luminosities probe somewhat different timescales of the star formation -- 100~Myr versus 200~Myr 
\citep{kennrev}. By obtaining such results (Table~\ref{tbl_uvres}) we may suspect that in the ring of NGC~1543
the star formation proceeds with a roughly constant rate for several hundreds Myr, while in NGC~1533 it may fade out.

\section{Discussion}

\subsection{The structure of NGC~1533 and NGC~1543.}

\begin{figure*}
\begin{tabular}{c c}
 \includegraphics[width=0.45\textwidth]{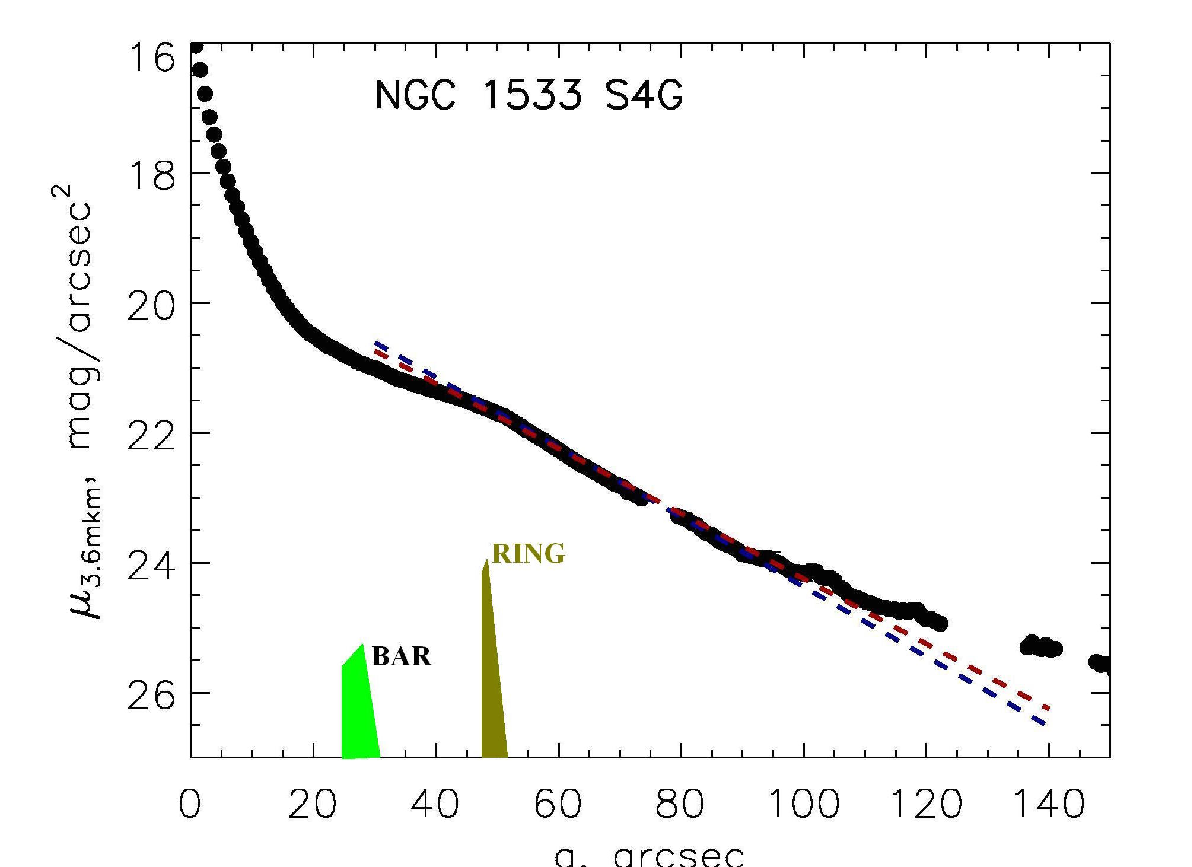} &
 \includegraphics[width=0.45\textwidth]{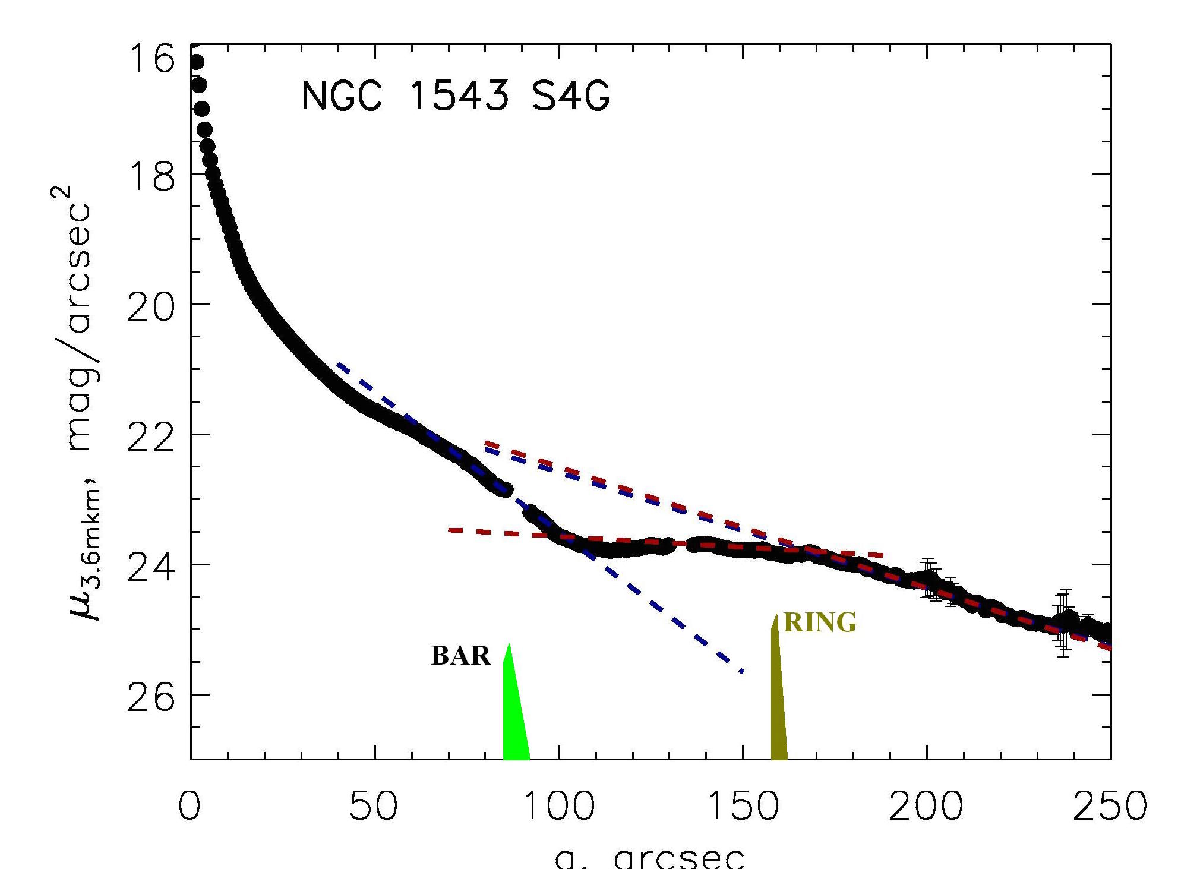} \\
\end{tabular}
\caption{The azimuthally-averaged surface brightness profiles in the 3.6$\mu$m band calculated by using the images 
from the S4G photometric survey \citep{sheth10}, for NGC~1533 ({\it left}) and NGC~1543 ({\it right}). 
The blue dashed lines are our exponential fits for the outer stellar discs
(and for the inner disc in NGC~1543) while the red dashed lines are exponential fits by \citet{Laine_S4G} 
explored and published in the frame of S4G survey.
}\label{profs}
\end{figure*}

Both galaxies are certainly classified as SB0 \citep[e.g.][]{south_rings}, so by definition their large-scale structures include
a bulge, a disc, and a bar. However, there were also many attempts to refine this simple description, and listing of additional
features for NGC~1533 and NGC~1543 differed when various investigators analysed images in different photometric bands. Both 
galaxies have been observed in the NIR photometric survey S4G with the space telescope Spitzer \citep{sheth10} so to be
concrete we have calculated and show their azimuthally-averaged surface brightness profiles in the 3.6$\mu$m band in Fig.~\ref{profs}.
Since both galaxies are seen nearly face-on there are no problems with flux averaging over an azimuthal angle at any fixed radius.  

NGC~1533 was classified as a ring galaxy as early as in 1995 \citep{south_rings} 
though its ring is not prominent too much (Fig.~\ref{images}). The size of its large-scale bar determined from 
the isophote analysis basing on position of the local isophote ellipticity maximum was estimated more than once: 
the estimates vary from $a_b=24.3^{\prime \prime}$ in the Carnegie-Irvine survey \citep{Carnegie-Irvine} to
$a_b=31.9^{\prime \prime}$ in the NIRS0S survey \citep{laurikainen_2006}.
The ring may thus be classified as a resonance structure because its radius, $50^{\prime \prime}$ \citep{ARRAKIS}, is larger 
by a factor of 1.5--2.0 than the radius of the bar that lies within the typical relation between the radius of the bar and 
the radius of its outer Lindblad resonance \citep{Buta2017}. About the bulge of NGC~1533 there were different opinions: 
some photometrists found the low Sersic parameter for it, $n=1.5$ \citep{laurikainen_2006} and treated it as a pseudobulge, 
but in the Carnegie-Irvine survey $n=2.39$ was found \citep{Carnegie_Irvine_2019} so the bulge can be classified as a classic one. Curiously, despite the difference in the decomposition parameters, the total contribution of the bulge into the galaxy luminosity 
was found to be $\sim 30\%$ in all surveys. The disc of NGC~1533 described as of II.i-type by \citet{Laine_S4G} 
in the frame of the S4G survey, is meantime a regular exponential disc with a typical scalelength of $21.3^{\prime \prime}$ 
\citep{Laine_S4G} or  $25.3^{\prime \prime}$ \citep{laurikainen_2006} or $20.2^{\prime \prime}$ (our fit in Fig.~\ref{profs}), 
hence in general around 2~kpc. At outer radii, at $R>100^{\prime \prime}$, or farther than 10~kpc from the center, 
some brightness excess is seen in Fig.~\ref{profs}, {\it left}, which may be treated as an outer stellar
disc of low surface brightness or as a stellar halo \citep{Cattapan2019} -- a remnant of minor merger. And it is just
to this outer structure that the orientation angles of the galaxy plane in Table~\ref{tbl_lit} refer -- they are valid for $R>120^{\prime \prime}$.

NGC~1543 looks quite differently: its ring is detached from the inner disc, and the radius of the ring is enormous, $160^{\prime \prime}$
\citep{laurikainen_2011}, or 15~kpc. But its bar is also large: the radius of the large-scale bar in NGC~1543 is estimated as
$84^{\prime \prime}$ (Carnegie-Irvine) to $92^{\prime \prime}$ \citep{Erwin2004}. Again the ratio of the ring-to-bar radii is about 2,
so the ring {\it may} have resonance nature. Inside the large bar one more nuclear bar is observed, with the radius of
$11^{\prime \prime}$ \citep{Erwin2004,laurikainen_2011} and with different orientation with respect to the large bar. The bulge is classical
and compact, contributing also only 30\%\ into the total luminosity \citep{Carnegie_Irvine_2019}. The disc surface brightness profile 
(Fig.~\ref{profs}, {\it right}) is even more typical for the 'truncated' class (II-type) of discs than that in NGC~1533, 
and it is really difficult to explain its whole
shape unambiguously: is it a brightness gap around $R\sim 100^{\prime \prime}$ or is it a very broad ring at $R > 150^{\prime \prime}$?
Several exponential segments can be fitted into the surface brightness profile over various radial ranges (Fig.~\ref{profs}, {\it right});
however we agree with \citet{Laine_S4G} that well beyond the ring, not only the stellar one with $R_{ring}=160^{\prime \prime}$, but
also beyond the ionised-gas starforming one at $R_{ring}=170^{\prime \prime}-190^{\prime \prime}$, the surface brightness profile 
can be well fitted with an exponential law assuming the radial scalelength of $1^{\prime}$, or about 5.5~kpc. 
Hence though the ring of NGC~1543 is a true outer ring, it is still embedded into the outer old stellar disc, very extended
and rather low-surface-brightness one.

\subsection{Large neutral hydrogen rings -- even more extended than the starforming ones}

Star formation in the rings must be fed by cold gas. Search for cold neutral hydrogen was undertaken in the 21~cm line for
NGC~1533 and NGC~1543, and rather large amounts of the gas were found (Table~\ref{tbl_lit}). However all this gas concentrates
in the outer parts of the galaxies. We can check if this gas has been accreted recently by considering its kinematics in
comparison with the kinematics of their stellar discs. Indeed, in early-type galaxies the kinematics of the HI discs -- and the
orientation of their planes -- is often decoupled from that of galactic stellar discs betraying recent cold-gas origin \citep{atlas3d_26}.

\citet{RyanWeber2004} presented a HI map for NGC~1533 obtained with the ATCA providing spatial resolution of about $30^{\prime \prime}$.
The gas is detected at distances from $2^{\prime}$ to $11.7^{\prime}$ off the optical center of NGC~1533 so well beyond the ring but
within the outer stellar disc of the galaxy. The shape of the gas distribution is an elliptical unclosed ring with the orientation
of the major axis (line of nodes?) at $PA\sim 120^{\circ}$, or close to the orientation of the outer stellar-disc line of nodes  
(Table~\ref{tbl_lit}).
However while the stellar disc looks almost circular, implying the inclination of $i\sim 20^{\circ}$, the ellipticity of the
HI ring requires  $i = 58^{\circ}$ if we suggest that the ring is intrinsically round and looks elliptical in projection due to
its inclination to our line of sight \citep{Murugeshan2023}. However, such high inclination of the gaseous disc means its asymptotic 
rotation velocity of only 130~\kms\ \citep{Murugeshan2023} that is inconsistent with the Tully-Fisher relation which we have
taken from \citet{Lelli_TF} to inspect the position of NGC~1533 on it with the galaxy baryonic mass derived from the data in Table~\ref{tbl_lit}.
More probably, the HI ring around NGC~1533 is intrinsically elliptical and unrelaxed being formed only recently due to minor merger
\citep{RyanWeber2004} having fallen from a highly inclined orbit. However in the tidal gravitational field of NGC~1533 
the accreted gas is able to inflow and to settle onto the resonance radius of the large bar by feeding then a local 
star formation burst in the ring. The velocities of the ionised gas in the ring (Fig.~\ref{vel1533}) are consistent 
with the HI velocities in this direction  \citep{Murugeshan2023}. Interestingly, the star formation site in the ring,
and one more UV-clump in the inner part of NGC~1533, are localised near the minor axis of the outer HI ring,
where the cold gas comes closer to the galaxy.

The HI ring in NGC~1543 was mapped also at the ATCA with the same resolution of about $30^{\prime \prime}$ and was presented in the paper by
\citet{Murugeshan2019}. But this time the HI ring looks round, has a radius of about $3^{\prime}$ and coincides exactly 
with the emission-line ring and UV-ring. Much more modest mass of the neutral hydrogen in NGC~1543 with respect to the HI mass 
in NGC~1533 (Table~\ref{tbl_lit}) can meantime be completely involved into the star formation process in the ring. 
The HI ring rotates, and rotates circularly at the first glance. By applying a popular tilted-ring software 3DBarolo \citep{3dbarolo}
for 3D fitting of the HI emission, \citet{Murugeshan2019} have found the parameters of the spatial orientation of the gaseous disc:
$PA_0 =134^{\circ}$ and $i=24^{\circ}$. While the inclination of the gaseous disc is consistent with that
of the stellar disc (Table~\ref{tbl_lit}), the line of nodes is strongly turned. Moreover, by comparing 
the line-of-sight velocity profiles obtained by us in two different position angles, we have found in Section~3 
that the kinematical line of nodes of the inner stellar disc is closer to $PA =170^{\circ}$ than to $PA =148^{\circ}$ 
being consistent with the orientation of the photometric major axis $PA =9^{\circ}$ and inconsistent 
with the line of nodes of the outer gaseous ring. These facts give an argument in favour of the inclined orientation
of the gaseous ring with respect to the plane of the stellar disc. Then we come to a conclusion that the cold gas feeding the
star formation in the NGC~1543 ring is also of external origin: it has been accreted and it conserves the orbital momentum of
the accretion source. 

\subsection{The origin of starforming rings in NGC~1533 and NGC~1543 and their evolution}

Usually classification of outer rings in disc galaxies divides them into resonance rings and accreted rings (plus
rather rare impact rings) -- see e.g. the review by \citet{Buta_Combes_1996}. However the case of NGC~1533 and NGC~1543
requires to introduce a composite type: the radii of the rings may be determined by the location of their bar resonances,
but the cold gas feeding their star formation comes evidently from outside. Not only the extended HI rings/discs demonstrate
quite independent, probably initial orbital kinematics, but also the line-of-sight velocities of the compact HII-regions 
in the rings, and beyond the ring of NGC~1533 \citep{RyanWeber2004}, match the kinematics of the outer HI structures 
while are decoupled from the kinematics of the inner galactic discs. Some subtle difference in the SFR estimates within
the rings from their FUV- and NUV-fluxes in NGC~1533 and NGC~1543 may indicate the stable dynamics of the gaseous ring
in the NGC~1543 and recent tidal disturbance of the outer gaseous ring in NGC~1533 resulting in brief gas inflow toward 
the resonance radius over the limited ring arc to the north from the galaxy center. Hence the local enhancement of star
formation rate in the northern part of the NGC~1533 ring may be caused by external impact. The star formation enhancement
in the south-eastern and north-western segments of the NGC~1543 ring cannot be explained by brief external impact
because all the gas is already concentrated in the ring. Let us note that the location of the star formation sites in
the ring is close to the orientation of the gaseous-disc line of nodes. If we suggest that the equatorial plane of the
NGC~1543 gravitational potential is seen almost face-on, perhaps, star formation sites mark the places where the
gaseous disc is crossing the equatorial plane of the NGC~1543 stellar disc (and its potential well respectively),
and so this local star formation is rather stable. However its time scale must also be rather brief, less than $10^9$ years 
-- as it is brief in particular in inner resonance rings \citep{kk04}, -- because we observe quasi-solar
metallicity of the ionised gas in the HII-regions in the ring. The gas metallicity in the outer starforming rings of
lenticular galaxies is almost always quasi-solar -- more exactly, $\langle \mbox{[M/H]}\rangle =-0.15$ 
independently on ring radius or galaxy luminosity \citep{s0_fp,Proshina2019}. We relate this homogeneity with the final stage
of chemical evolution in a starburst when the gas metallicity reaches its asymptotic value \citep[e.g.][]{ascasibar}.
Then to explain why we observe always just the final stages we must suggest a very brief and effective process
of gas re-processing into stars in the galactic outer rings.

\section*{Acknowledgements} 
The analysis of inclined gaseous discs in lenticular galaxies is supported by the grant of the Russian Science Foundation
no. 22-12-00080.
The work is based on the data obtained at the Southern African Large
Telescope and on the public data of GALEX (http://galex.stsci.edu/GR6/) and WISE surveys.
The NASA GALEX mission data were taken from the Mikulski Archive for Space
Telescopes (MAST). The WISE data exploited by us were retrieved from the NASA/IPAC Infrared Science Archive,
which is operated by the Jet Propulsion Laboratory, California Institute of Technology,
under contract with the National Aeronautics and Space Administration.
We acknowledge the usage of the HyperLeda database \url{http://leda.univ-lyon1.fr} \citep{hyperleda} 
and NASA Extragalactic database \url{http://ned.ipac.caltech.edu}. 
We acknowledge the usage of the Legacy Survey data \url{https://www.legacysurvey.org} \citep{legacysurvey}.
The Legacy Surveys consist of three individual and complementary projects: the Dark Energy Camera Legacy 
Survey (DECaLS; Proposal ID \#2014B-0404; PIs: David Schlegel and Arjun Dey), 
the Beijing-Arizona Sky Survey (BASS; NOAO Prop. ID \#2015A-0801; PIs: Zhou Xu and Xiaohui Fan), 
and the Mayall z-band Legacy Survey (MzLS; Prop. ID \#2016A-0453; PI: Arjun Dey). 
DECaLS, BASS and MzLS together include data obtained, respectively, at the Blanco telescope, 
Cerro Tololo Inter-American Observatory, NSF's NOIRLab; the Bok telescope, Steward Observatory,
University of Arizona; and the Mayall telescope, Kitt Peak National Observatory, NOIRLab. 
Pipeline processing and analyses of the data were supported by NOIRLab and 
the Lawrence Berkeley National Laboratory (LBNL). The Legacy Surveys project is honored 
to be permitted to conduct astronomical research on Iolkam Du'ag (Kitt Peak),
a mountain with particular significance to the Tohono O'odham Nation.

\section*{Data availability}

The data underlying this article will be shared on reasonable request to the corresponding author.

\bibliographystyle{mnras}                                                                                              
\bibliography{saltdorado_kniazev}

\appendix

\section{The spectra of the main structure components in NGC~1533 and NGC~1543}

Below we present some spectra which have been used for the analysis of the stellar population properties
in Section~5 and for deriving the ionised-gas excitation and oxygen abundance in Section~4.

\begin{figure*}
	\centering
	\includegraphics[width=0.9\textwidth]{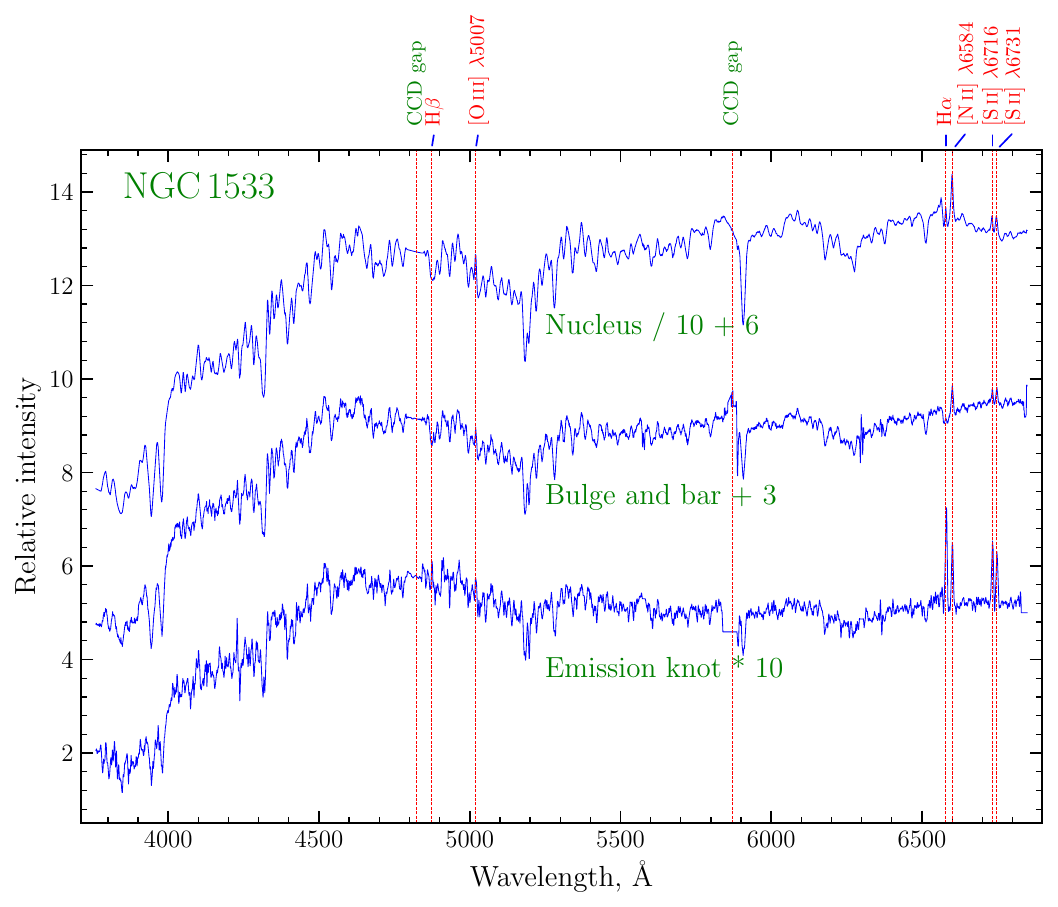}
	\caption{NGC~1533: The SALT/RSS spectra of the main structure components of the galaxy. The borders of binning
        to integrate the spectra for the nucleus and for the bulge$+$bar are those mentioned in the caption of Fig.~\ref{lick1533}.
         The positions of the main emission lines and of the gaps in the RSS CCD mosaic are marked. The abscissa units (wavelengths)
         are reduced to the restframe. The ordinate units are
         relative intensities; for the purpose of the visualisation, the spectrum of the ring knot is multiplied by a factor of 10,
         and the spectra of the nucleus (divided by 10) and of the bulge$+$bar are shifted up.}
	\label{app_sp1533}
\end{figure*}

\begin{figure*}
	\centering
	\includegraphics[width=0.9\textwidth]{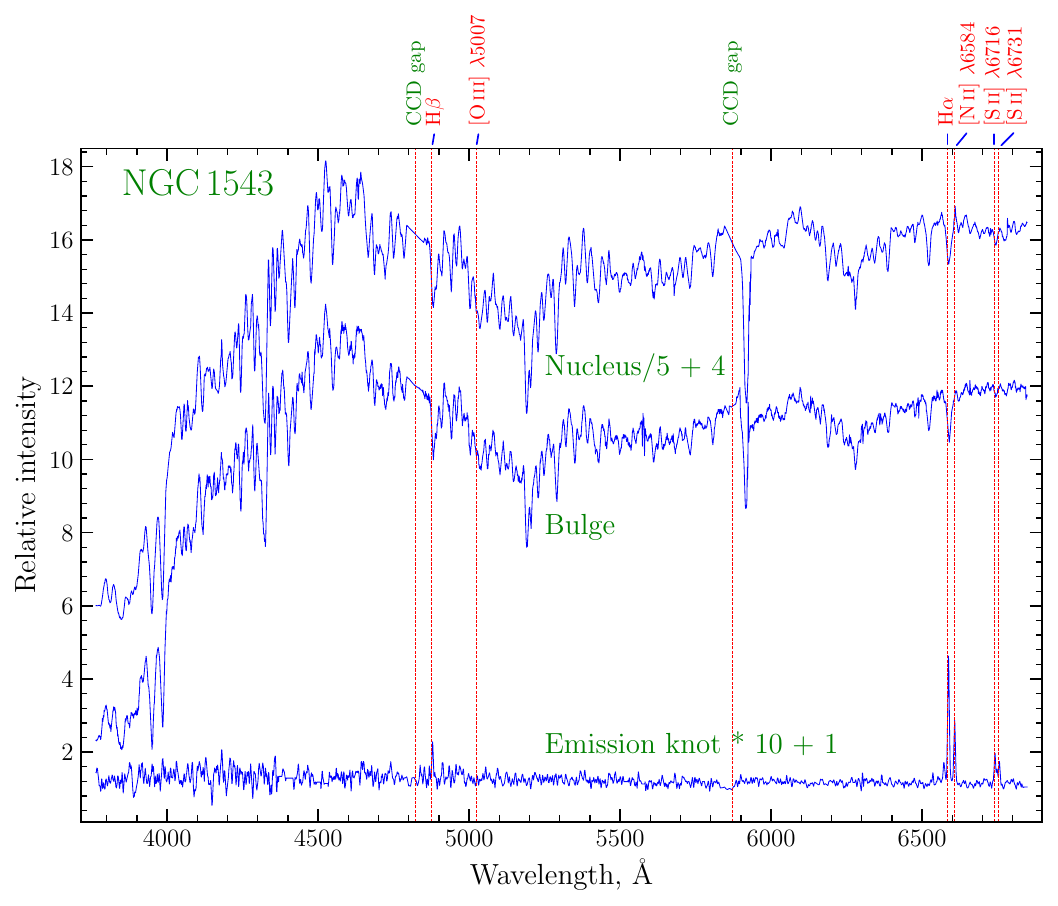}
	\caption{NGC~1543: The SALT/RSS spectra of the main structure components of the galaxy. The borders of binning
        to integrate the spectra for the nucleus and for the bulge are those mentioned in the caption of Fig.~\ref{lick1543}.
         The positions of the main emission lines and of the gaps in the RSS CCD mosaic are marked. The abscissa units (wavelengths)
         are reduced to the restframe. The ordinate units are
         relative intensities; for the purpose of the visualisation, the spectrum of the ring knot is multiplied by a factor of 10,
         and the spectrum of the nucleus is divided by 5, and both are shifted up.}
	\label{app_sp1543}
\end{figure*}


\bsp	

\label{lastpage}

\end{document}